\documentclass{emulateapj}
\slugcomment{Accepted to \apj: April 6, 2015} 
\usepackage{natbib}
\usepackage{amsmath}
\usepackage{booktabs}
\defcitealias{GO2011}{GO11}
\defcitealias{GO2009}{GO09}

\newcommand{\di}{\mathrm{d}}

\begin{document}
\title{Prestellar Core Formation, Evolution, and Accretion from Gravitational
Fragmentation in Turbulent Converging Flows}
\author{Munan Gong and Eve C. Ostriker}
\affil{Department of Astrophysical Sciences, Princeton University,
Princeton, New Jersey 08544, USA}
\email{munan@princeton.edu, eco@astro.princeton.edu}

\begin{abstract}
    We investigate prestellar core formation and accretion 
    based on three-dimensional hydrodynamic simulations.
    Our simulations represent
    local $\sim 1$pc regions within giant molecular clouds where a
    supersonic turbulent flow converges, triggering star formation in the
    post-shock layer. We include turbulence and self-gravity,
    applying sink particle techniques,
    and explore a range of inflow Mach number ${\cal M}=2-16$. 
    Two sets of cores are identified and compared: 
    $t_1$-cores are identified of a time snapshot in each simulation,
    representing dense structures in a single cloud map;
    $t_\mathrm{coll}$-cores are identified at their individual time of
    collapse, representing the initial mass reservoir for accretion.
    We find that cores and filaments form and evolve at the same time. At the
    stage of core collapse, there is a well-defined, converged characteristic
    mass for isothermal fragmentation that is comparable to the critical Bonner-Ebert
    mass at the post-shock pressure. The core mass functions (CMFs) of 
    $t_\mathrm{coll}$-cores show a deficit of high-mass cores
    ($\gtrsim 7M_\sun$) compared to the observed stellar initial mass
    function (IMF). However, the CMFs of $t_1$-cores are similar to the observed
    CMFs and include many low-mass cores that are gravitationally stable. 
    The difference between $t_1$-cores and $t_\mathrm{coll}$-cores suggests
    that the full sample from observed CMFs
    may not evolve into protostars.
    Individual sink particles accrete at a roughly constant rate throughout the
    simulations, gaining one $t_\mathrm{coll}$-core mass per free-fall
    time even after the initial mass reservoir is accreted. High-mass sinks
    gain proportionally more mass at late times than low-mass sinks.
    There are outbursts in accretion rates, resulting
    from clumpy density structures falling into the sinks.
\end{abstract}


\section{Introduction}
Dense molecular cores form in the last step of giant molecular cloud (GMC)
fragmentation and are the cradles of single stars or close binary systems
\citep{MO2007, Andre2014}.
These cores provide the initial mass reservoir and set the local environment of
star formation. Understanding core formation and the core mass function
(CMF) is important for characterizing the initial conditions of star and planet
formation, and explaining the origin of the stellar initial mass function
\citep[IMF, see review of][]{Offner2014}. 

Cores emerge in dense regions of GMCs, the dynamics of which are
governed by turbulence, gravity, and magnetic fields \citep{Dobbs2014}. 
Although supersonic turbulence provides support for GMCs on global scales, 
it compresses gas to high densities in local regions where gravitational collapse can
rapidly occur \citep{BP2007, MO2007}. Strong turbulence in GMCs can also accelerate
ambipolar diffusion and aid in the creation of magnetically supercritical
cores \citep{NL2005, NL2008, KB2008, KB2011, CO2012}.
However, in the situation where the large-scale GMC is magnetically
supercritical, simulations show that anisotropic core formation via magnetic
field-aligned flows results in similar core masses to the unmagnetized case
\citep{CO2014}. In this paper, we focus on core formation and
subsequent collapse and accretion using turbulent, unmagnetized simulations.

Cores are at the bottom of the hierarchy in a GMC, and are often
associated with dense filaments seen in dust continuum and molecular lines. 
Observations show many pressure-confined gravitationally stable cores within
clouds (and filaments), which may later become gravitationally unstable
\citep{Lada2008, Andre2010, Schisano2014}.
High resolution observations by {\sl Herschel} Space Observatory indicate
that gravitationally bound cores are mainly embedded along thermally super-critical
filaments \citep{Andre2010, Molinari2010, Schisano2014}, i.e., those with mass
per unit length exceeding $2c_s^2/G$ \citep{Ostriker1964}. 
This is in apparent agreement with the theoretical conclusion that 
self-gravitating filaments are unstable to longitudinal perturbations 
\citep{Nagasawa1987, IM1992, IM1997}, which would allow them to fragment into
dense cores. However, the most unstable scale is several times the filament
diameter, which would produce cores more massive than observed.
Also, the two-step scenario in which filaments first form
and then subsequently fragment into cores can be altered with the presence
of large-amplitude turbulence, because the nonlinear perturbations produced by
turbulence allow structures at many scales to grow simultaneously.
With numerical
simulations, it is possible to follow the temporal evolution of developing
structures, and simulations with turbulence show that filaments and cores
develop at the same time rather than in sequence \citep[see also \S
\ref{section:filaments} in this paper]{GO2011, Loo2014, CO2014}.

Observationally, cores are divided into two categories: starless cores,
and protostellar cores/Class 0 objects \citep{diFrancesco2007, Andre2009}.
Among the starless cores, a further classification is of prestellar cores,
which are strongly gravitationally bound and centrally concentrated;
the balance of other starless cores
are the opposite: they appear to be largely pressure-confined rather than
strongly gravitationally bound,
and are not strongly centrally concentrated. 
Protostellar cores/Class 0 objects are gravitationally bound and contain 
embedded protostars/young stellar objects (YSOs). 
It is generally thought that these categories represent cores at different
evolutionary stages: pressure-confined starless cores gain mass and turn into gravitationally bound
prestellar cores, which finally collapse and become protostellar cores. 
However, in practice,
prestellar cores are often not distinguished from other starless cores in
observations. Some starless cores may
even be transient structures that will never collapse to form protostars.
Numerical simulation made it possible to check whether observed
pressure-confined cores are likely to become unstable at a later time.

Many theoretical studies have investigated individual core evolution, beginning with
the semi-analytical work of \citet{Larson1969}, \citet{Penston1969}, and \citet{Shu1977}.
Based on isothermal simulations with supersonic converging flows, 
\citet{GO2009} identified four stages of core evolution:
core building, core collapse, envelope infall, and late accretion. 
In the core building stage, the core gains mass from its environment, gradually
becoming more centrally concentrated. For the spherical case, shock-bounded
cores collapse when the central-to-edge pressure ratio is similar to that of a
critical Bonnor-Ebert equilibrium sphere \citep{Ebert1955, Bonnor1956}, which
has mass
\begin{align}\label{eq:M_BE_crit}
\begin{split}
    M_{BE}&=1.2 \frac{c_s^4}{(G^3P_\mathrm{edge})^{1/2}} = 1.9
    \frac{c_s^3}{(G^3\bar{\rho})^{1/2}}\\
    &= 2.3 M_\sun \left( \frac{\bar{n}_H}{10^4 \mathrm{cm}^{-3}}
    \right)^{-1/2} \left( \frac{T}{10\mathrm{K}} \right)^{3/2}.
\end{split}
\end{align}
Here $P_\mathrm{edge}$ is external pressure at the edge of the core;
$\bar{\rho}$ is the mean core density; $T$ the core's temperature; and
$c_s=(kT/\mu)^{1/2}$ is the internal sound speed in the core. 
After the core collapses, the internal density structure approaches
the well-known Larson-Penston solution \citep[hereafter LP]{Larson1969, Penston1969}
\begin{equation}\label{eq:rho_LP}
    \rho_{\mathrm{LP}}(r) = \frac{8.86 c_s^2}{4\pi G r^2},
\end{equation}
where $r$ is the distance from the core center.
This asymptotic state is reached regardless of the dynamical evolution leading
up to collapse \citep[see discussion and references in][hereafter 
\citetalias{GO2009} and \citetalias{GO2011}]{GO2009, GO2011}; in particular,
cores formed by supersonic flows approach the LP solution during collapse. 
Collapse leads to creation of an opaque core and then 
a protostar at the center, with a surrounding rotating disk \citep{Shu1987}. 
Subsequently, the dense envelope falls onto the protostar via an inside-out
gravitationally-induced
rarefaction wave \citep{Shu1977, Hunter1977}. \citetalias{GO2009} define the late
accretion stage as the period after the rarefaction wave has propagated beyond
the region where the LP profile holds, and most of the original core mass has
fallen into the center (the protostar/disk). The characteristic timescale for
the core collapse and the envelope infall stages is the free-fall time
\begin{equation}\label{eq:tff}
    t_\mathrm{ff} = \left( \frac{3\pi}{32G\bar{\rho}} \right)^{1/2}
    = 4.3\times 10^5\mathrm{yr}
    \left( \frac{\bar{n}_H}{10^4 \mathrm{cm}^{-3}} \right)^{-1/2}.
\end{equation}
This is similar to the lifetime of prestellar and protostellar cores
in observations \citep{Andre2014}. 

Empirical measurement of CMFs from extinction and continuum mapping 
suggest the mass distribution has a similar
shape to the IMF, but with the characteristic core mass shifted upward
by a factor of $\sim 3$ \citep[e.g.,][]{Alves2007, NW2007, Konyves2010}.
However, observed CMFs often include transient structures that may not collapse later,
and the core masses obtained from observations can be dependent
on the specific core finding algorithm applied (e.g., {\sl gaussclumps}
\citep{SG1990}; {\sl clumpfind} \citep{Williams1994}; {\sl getsources}
\citep{Menshchikov2010}).
Although core identification based directly on surface density maps is somewhat
ambiguous due to the arbitrary definition of the core boundary, use of the gravitational
potential removes some of this ambiguity \citep{Smith2009, GO2011}.
Analytical theories of the CMF \citep[see review of][]{Offner2014} focus
on gravitationally bound cores. In this work, we identify two sets of cores:
The first set is comparable to structures seen in a GMC at a given instant in time,
and includes a range of evolutionary stages (both weakly and strongly bound).
The second set represents cores
seen at a fixed evolutionary state, the time of protostar formation. We address
the differences between the two sets, and implications for interpretation of
observations.

Even if every observed core were to collapse, evolution after core formation could
further change the relation between the CMF and IMF. At a minimum, fragmentation
into binaries and multiples due to rapid rotation, and mass loss due to
protostellar outflows, complicates any mapping from the CMF to the IMF. In
addition, a concern in the theoretical literature has been that 
there may be no preferred scale for isothermal fragmentation such that
gravitational fragmentation in GMCs must depend crucially on
non-isothermal effects \citep{Martel2006, Krumholz2014}. 
This concern arises in part because an isothermal filament with mass per unit
length greater than $2c_s^2/G$ can collapse towards its axis without
fragmenting, leading to infinite density \citep{IM1997}, and in part because 
homologous collapse of a sphere would lead to ever smaller Jeans masses and possible
sub-fragmentation until stopped by non-isothermal effects \citep{Hoyle1953}.

In fact, core collapse is non-homologous, such that the hierarchical
fragmentation envisioned by \citet{Hoyle1953} does not occur. However, disks
that form around protostars are known to undergo excessive fragmentation and
overproduce brown dwarfs if stellar heating is not included in simulations
\citep{Bate2009}. Here, we argue that the problem of excessive disk
fragmentation in isothermal models can be distinguished form whether there is a
well-defined CMF from gravoturbulent fragmentation of 
an isothermal cloud. We shall show that
the characteristic core mass converges with increasing resolution (and 
threshold of sink particle creation) in our simulations. 
In reality, of course, GMCs and cores are not isothermal, but have temperature
variations between $\sim 5-10$K, with observationally inferred temperatures
dropping toward the center of the cores \citep[e.g.,][]{Marsh2014,Roy2014}.
However, the isothermal approximation is an adequate first limit from
cloud to core scales. Radiation feedback and significant non-isothermal effects become
important subsequent to core collapse, in the accretion disks around protostars.         

In this paper, we present our results on prestellar core formation and  collapse 
based on a large suite of three-dimensional numerical simulations. 
Our simulations model a local ($\sim 1$pc) region of GMCs where a supersonic
turbulent flow converges, inducing star formation in the post-shock layer. 
We adopt the same model of a supersonic and isothermal
converging flow with turbulent perturbations as in \citetalias{GO2011}.
As in \citetalias{GO2011}, we see filamentary structures form at the
same time as dense cores. While the models of \citetalias{GO2011} were halted
when the first core collapsed, here we apply
the sink particle implementation of \citet{GO2013} to trace the evolution at
later times. We also explore a wider range of inflow Mach numbers (${\cal M}=2-16$).
Using simulations with varying numerical resolution, we show that the
characteristic core mass at the time of collapse is well defined. 
We measure the statistical properties of cores,
especially the CMF, and compare to the observed CMF and IMF. We are also
able to trace the accretion rate of individual sink particles, and explore its
time variation and dependence on core masses.

The outline of this paper is as follows. We describe our numerical methods and
model parameters in \S \ref{section:methods}. In \S \ref{section:results}, we
present our results of overall evolution (\S \ref{section:evolution}),
convergence of isothermal fragmentation (\S \ref{section:convergence}), core
properties and CMF (\S \ref{section:core}), and sink particle accretion (\S
\ref{section:sink}). Finally \S \ref{section:summary} summarizes our
conclusions.

\section{Methods and Parameters}\label{section:methods}
\subsection{Numerical Methods}

\subsubsection{Basic Equations and Algorithms}
The numerical simulations presented in this paper are conducted with the 
{\sl Athena} code \citep{Stone2008, SG2009},
on a three dimensional (3D) Cartesian grid, using the van Leer (MUSCL-Hancock
type) integrator, the HLLC Riemann solver, and
second-order spatial reconstruction. The code solves for the self-gravity in planar geometry,
with open boundary condition in the vertical ($\hat{z}$) direction, and periodic
boundary conditions in the horizontal ($\hat{x}$ and $\hat{y}$) directions, 
using the fast Fourier transformation (FFT) method developed by \citet{KO2009}.
Sink particles as implemented by \citet{GO2013} are included, and allow us to
track the system evolution after individual cores collapse. 

The equations solved for the gas are the conservation of mass and momentum:
\begin{equation}\label{eq:continuity}
    \frac{\partial \rho}{\partial t} + \nabla \cdot (\rho \mathbf{v}) = 0, 
\end{equation}
\begin{equation}
    \frac{\partial \mathbf{v}}{\partial t} +  \mathbf{v} \cdot \nabla
    \mathbf{v} = - \frac{\nabla P}{\rho} - \nabla (\Phi + \Phi_{\mathrm{sp}}),
\end{equation}
and the Poisson equation:
\begin{equation}
    \nabla^2 \Phi = 4\pi G \rho,
\end{equation}
where $\rho$ is the density, $\mathbf{v}$ the velocity, $P$ the pressure, 
$\Phi$ the gravitational potential of the
gas, and $\Phi_{\mathrm{sp}}$ the gravitational potential associated with the
sink particles. For simplicity, we adopt an isothermal equation of state for the gas:
\begin{equation}
    P = c_s^2 \rho,
\end{equation}
where $c_s = \sqrt{kT/(2.3 m_H)} = 0.19\mathrm{km/s}~(T/10\mathrm{K})^{1/2}$ 
is the isothermal sound speed of molecular gas with solar metallicity;
$T$ the temperature; and $m_H$ the mass of the hydrogen atom.

The sink particle implementation of \citet{GO2013} is briefly
summarized here. Our sink particle creation criteria require a local potential
minimum and the density threshold $\rho_{\mathrm{thr}}$ 
set by the Larson-Penston profile in Equation
(\ref{eq:rho_LP}) at a distance $\Delta x/2$ from the center:
\begin{equation}\label{eq:rho_thr}
    \rho_{\mathrm{thr}} = \rho_{\mathrm{LP}}(0.5 \Delta x) = 
    \frac{8.86}{\pi} \frac{c_s^2}{G\Delta x};
\end{equation}
here $\Delta x$ is the simulation cell size. Previously, 
\S 4.3 of \citet{GO2013} compared
different criteria for sink particle creation \citep[cf.][]{Bate1995,
Krumholz2004, Federrath2010, PN2011}, including the Truelove density
threshold \citep{Truelove1997}, additional checks for converging flow and a 
gravitationally bound state, and found no differences in sink particle
evolution. In our algorithm,
there is a moving control volume $(3\Delta x)^3$ centered on the cell 
containing each sink particle that acts similar to the ghost
zones of the simulation domain. At every time step, the density and momentum of
the cells inside each control volume are reset by extrapolation of the
surrounding non-sink zones.
The gas mass and momentum accretion onto each sink particle are
calculated based on the sum over all fluxes returned by the Riemann solver
at the outer surfaces of the sink control volume. 
Flows onto sink particles due to shifting of the control volume are also
included.
Sink particle positions and velocities are advanced in time
via a leapfrog integrator. The gravitational potential from and the forces on sink
particles are computed using a particle mesh method with a TSC kernal
\citep{HE1981}. Sink particles merge if their control volumes overlap.

\subsubsection{Problem Specification and Units\label{section:units}}
We used the prescription of converging supersonic flow with turbulent
perturbations following \citetalias{GO2011}
and \citet{GO2013}. The models represent a
localized ($\sim \mathrm{pc}$ scale) region in a turbulent GMC 
where the large scale velocity field 
converges and pre-stellar cores form in the post-shock dense gas. 

The model setup consists of supersonic flow augmented with turbulent velocity
perturbations, converging to the mid-plane from $+z$ and $-z$ directions at an
average velocity of $-{\cal M} c_s$ and $+{\cal M} c_s$. The inflowing gas has
the same spectrum of turbulent perturbations as the initial conditions (see below), i.e.,
the cloud ``outer scales" are treated as maintaining a fixed turbulent
amplitude. There is no explicit turbulent driving within the box, although
turbulence can be driven by shock instabilities.
In our parameter survey, the Mach
number ${\cal M}$ values are 2, 4, 8 and 16. The initial density is set equal to 
the code unit
$\rho_0$ (see below), as is the density of the converging flow from the $\hat{z}$
boundaries. The boundaries in $\hat{x}$ and $\hat{y}$ directions are periodic.

We apply velocity perturbations for both the whole domain initially and the
inflowing gas subsequently, following the prescription of \citetalias{GO2011}.
The scaling law of the velocity field represents the observed linewidth-size
relation in GMCs over a range of scales\footnote{Note, however, that at 
small scales where the turbulence is subsonic, it may follow a shallower Kolmogorov 
power law.} \citep{MO2007}:
\begin{equation}\label{eq:dv}
    \delta v(l) \propto l^{1/2}.
\end{equation}
However, the timescale for the turbulence to relax at the largest scales, 
attaining self-consistent density and velocity perturbations, is longer than the inflow
time from the boundaries. Thus, although the turbulence is not fully self-consistent, 
it serves to produce density perturbations that then evolve to collapse gravitationally.

Assuming the cloud-scale supersonic turbulence provides both
the kinetic support of the cloud against its self-gravity and
the large-scale converging flow with the virial parameter
$\alpha_\mathrm{vir}\sim 1$, \citetalias{GO2011} show 
(their Equation (43) and (44))
that the Mach number of the converging flow can be related to the cloud's size
$R_\mathrm{cloud}$ and Jeans length $L_J$ as ${\cal M} \sim R_\mathrm{cloud}/L_J$.
With the scaling of the velocity in Equation (\ref{eq:dv}), the
amplitude of velocity perturbations at the box scale $L$ 
can be written as $v_{1\mathrm{D}}(L) = ({\cal M} L/L_J)^{1/2} c_s$
\footnote{In a global model, for which $L\rightarrow2R_\mathrm{cloud}$, 
one can show that this
reduces to $V_\mathrm{1D}(2R_\mathrm{cloud})\sim {\cal M}c_s$.}.
This corresponds to the ``high-amplitude" 
perturbation simulations in \citetalias{GO2011}. In this paper, we adopt 10\%
of the above values for velocity perturbations, corresponding to the
``low-amplitude" case in \citetalias{GO2011}. 
Thus, the largest velocity perturbation in a simulation is limited by
\begin{equation}\label{eq:dv_1D}
v_{1\mathrm{D}}(L) = 0.1({\cal M} L/L_J)^{1/2} c_s,
\end{equation}
where $L$ is the
horizontal ($\hat{x}$ and $\hat{y}$ direction) box size. 
Note that for the Mach numbers and box size we use, 
although the average inflow velocity is highly supersonic, the velocity perturbations 
within the domain are subsonic.

We choose to focus on the low-amplitude perturbation case for two
reasons: Firstly, we wish to concentrate on the local details of collapse produced by
converging flows. For realistic GMC conditions, ${\cal M}\gg 1$ and the
velocity perturbations at scales $\sim L_J$ are small compared to the inflow
motion ${\cal M}c_s$, which means a post-shock layer with simple geometry would
form; this situation is amenable to local simulations. However, for
completeness, we wish to explore a range of ${\cal M}$ values down to ${\cal
M}=2$, and  at low ${\cal M}$ the velocity perturbations at scales $\sim L_J$
would be comparable to inflow speeds. In this case,
there would no longer be a post-shock layer with simple geometry, so that 
global simulations would be more suitable than local. By reducing the
perturbation amplitude, we can use local models for all ${\cal M}$.
Secondly, we are limited
by computational power in practice, and the CPU time required for a simulation
run is increased in high amplitude cases. Also, bigger simulation boxes and
higher resolutions are required when the post-shock layer does not have a
simple geometry and shows more complicated dynamics, increasing the necessary number of
zones.

Nonetheless, it is important to address how the level of turbulence affects
the core formation process. We therefore conducted two simulations with
high-amplitude perturbations (see Section \ref{section:model_parameters} for
model parameters), and compare to the low-amplitude cases.

The code unit of density $\rho_0$ represents the average ambient density in a
GMC flowing into the region in which stars will form. For this reference density,
a characteristic spatial scale for gravitational instability is the Jeans length
\begin{equation}\label{eq:L_J}
    L_J \equiv c_s \left( \frac{\pi}{G\rho_0} \right)^{1/2}.
\end{equation}
The corresponding Jeans mass is
\begin{equation}\label{eq:M_J}
    M_J \equiv \rho_0 L_J^3 
        = c_s^3 \left( \frac{\pi^3}{G^3\rho_0} \right)^{1/2},
\end{equation}
and Jeans time is
\begin{equation}
    t_J \equiv \frac{L_J}{c_s} 
        = \left( \frac{\pi}{G\rho_0} \right)^{1/2} = 3.27t_{ff}(\rho_0).
\end{equation}
We use the Jeans length, mass, and time at the unperturbed density as the code
units for length, mass, and time: $L_0 = L_J$, $M_0 = M_J$ and $t_0 = t_J$.
In making comparison to observations, another useful quantity is 
the surface density integrated along the z-direction through the domain
\begin{equation}\label{eq:sigma}
    \Sigma = \int \rho(x,y,z)\di z 
           = \Sigma_0 \int \frac{\rho}{\rho_0} \frac{\di z}{L_J},
\end{equation}
where $\Sigma_0 \equiv \rho_0 L_J$.

To convert code units of length, mass and time to physical units, one may
choose an appropriate mean density, $\rho_0 = 1.4 n_{H,0} m_H$ (where $n_H =
2n_{H_2}$), and a
temperature $T$. This allows consideration of GMCs with a range of properties.
Many GMCs are observed to be self-gravitating, such that the virial parameter
$\alpha_\mathrm{vir}\equiv 5R_\mathrm{GMC}\sigma_{v, 1D}^2/(G M_\mathrm{GMC})$ 
is order-unity, where
$\sigma_{v, 1D}$ is the large-scale velocity dispersion along any direction.
Defining the density of the cloud as $\rho_0 \equiv M_\mathrm{GMC}/(\frac{4}{3}\pi
R_\mathrm{GMC}^3)$ and the surface density as 
$\Sigma_\mathrm{GMC}=4\rho_0 R_\mathrm{GMC}/3$ so that
$R_\mathrm{GMC}=5\sigma_{v,1D}^2/(\alpha_\mathrm{vir}\pi G\Sigma_\mathrm{GMC})$
, we can substitute
$\rho_0 = 3\pi \alpha_\mathrm{vir} G \Sigma_\mathrm{GMC}^2/ (20 \sigma_{v, 1D}^2)$
and $\sigma_{v, 1D} = {\cal M}c_s$ (assuming that the inflow Mach number ${\cal
M}$ is related to the large-scale GMC's turbulence level), to obtain 
\begin{equation}
L_0 = \left(\frac{20}{3\alpha_\mathrm{vir}}\right)^{1/2} {\cal M}\frac{c_s^2
}{G\Sigma_\mathrm{GMC}},
\end{equation}
\begin{equation}
M_0 = \pi \left(\frac{20}{3\alpha_\mathrm{vir}}\right)^{1/2} {\cal M}
\frac{c_s^4}{G^2\Sigma_\mathrm{GMC}},
\end{equation}
\begin{equation}
t_0 = \left(\frac{20}{3\alpha_\mathrm{vir}}\right)^{1/2} {\cal M}\frac{c_s
}{G\Sigma_\mathrm{GMC}},
\end{equation}
and
\begin{equation}
\Sigma_0 = \pi
\left(\frac{3\alpha_\mathrm{vir}}{20}\right)^{1/2} \frac{1}{{\cal M}} \Sigma_\mathrm{GMC}.
\end{equation}
These forms are useful because observations of GMCs measure surface density and Mach
number more directly than volume density. Many of the best-studied local dark
clouds and inner-galaxy GMCs
have surface density $\Sigma_\mathrm{GMC}\approx 200M_\sun/\mathrm{pc}^2$, although
clouds in more extreme environments such as the Galactic center and
ultra-luminous infrared galaxies (ULIRGs) have higher
$\Sigma_\mathrm{GMC}$ \citep{Dobbs2014}.  Table \ref{table:units}
lists the two different ways of translating code units to physical
parameters in molecular clouds.

\begin{table*}[htbp]
    \centering
    \caption{Code units and corresponding physical units \label{table:units}}
    \begin{tabular}{l  ll}
        \tableline
        \tableline
              $L_0$ 
           &$0.87\mathrm{pc}\left( \frac{n_{H,0}}{10^3 \mathrm{cm}^{-3}} \right)^{-\frac{1}{2}}
        \left( \frac{T}{10\mathrm{K}} \right)^{\frac{1}{2}}$  
            &$0.77\mathrm{pc} \left( \frac{{\cal M}}{10} \right)
            \left( \frac{\alpha_\mathrm{vir}}{2} \right)^{-\frac{1}{2}} 
            \left( \frac{T}{10\mathrm{K}} \right)
            \left( \frac{\Sigma_\mathrm{GMC}}{200M_\sun/\mathrm{pc}^2} \right)^{-1}$ \\
        $M_0$ 
          &$23 M_\sun \left( \frac{n_{H,0}}{10^3 \mathrm{cm}^{-3}} \right)^{-\frac{1}{2}}
            \left( \frac{T}{10\mathrm{K}} \right)^{\frac{3}{2}} $ 
          &$20 M_\sun \left( \frac{{\cal M}}{10} \right)
            \left( \frac{\alpha_\mathrm{vir}}{2} \right)^{-\frac{1}{2}}
            \left( \frac{T}{10\mathrm{K}} \right)^2
            \left( \frac{\Sigma_\mathrm{GMC}}{200M_\sun/\mathrm{pc}^2} \right)^{-1}$\\
        $t_0$ &$4.4 \mathrm{Myr} 
             \left( \frac{n_{H,0}}{10^3 \mathrm{cm}^{-3}} \right)^{-\frac{1}{2}} $ 
              &$4.0 \mathrm{Myr} \left( \frac{{\cal M}}{10} \right)
              \left( \frac{\alpha_\mathrm{vir}}{2} \right)^{-\frac{1}{2}}
              \left( \frac{T}{10\mathrm{K}} \right)^{\frac{1}{2}}
              \left( \frac{\Sigma_\mathrm{GMC}}{200M_\sun/\mathrm{pc}^2} \right)^{-1}$\\
        $\Sigma_0$ &$30M_\sun \mathrm{pc}^{-2} 
             \left( \frac{n_{H,0}}{10^3 \mathrm{cm}^{-3}} \right)^{\frac{1}{2}}
              \left( \frac{T}{10\mathrm{K}} \right)^{\frac{1}{2}}$
              &$34M_\sun \mathrm{pc}^{-2} 
              \left( \frac{{\cal M}}{10} \right)^{-1}
                  \left( \frac{\alpha_\mathrm{vir}}{2} \right)^{\frac{1}{2}}
                 \left( \frac{\Sigma_\mathrm{GMC}}{200M_\sun/\mathrm{pc}^2} \right) $\\
        \tableline
    \end{tabular}
    \tablecomments{Code units are given as a function of reference volume
        density $n_{H,0}$ (left) and based on a corresponding cloud surface density 
        $\Sigma_\mathrm{GMC}$ and Mach number ${\cal M}$ (right).
    }
\end{table*}

\subsubsection{Core Finding Method}
The cores in the simulations are identified 
using the {\sl GRID} core finding method developed by
\citetalias{GO2011}: the region belonging to a core is defined by the largest
closed contour of the gravitational potential field around the corresponding
local potential minimum that contains no other local minimum.
Within each core, we further define the bound core region as all the
material that has the sum of gravitational energy and 
thermal energy negative.\footnote{The gravitational and thermal energy density
    in each zone is $E_G=-\rho(\Phi_{\mathrm{max}}-\Phi)$, 
    where $\Phi_{\mathrm{max}}$ is the
gravitational potential at the core boundary, and $E_{\mathrm{th}}= 3/2\rho
c_s^2$. We have also defined kinetically bound cores with the volume integral
of $E_\mathrm{th}+E_K+E_G$ less than zero, 
where $E_K=(v_x^2+v_y^2+v_z^2)/2$ is the kinetic
energy of the gas. We found that most cores are mainly supported by
thermal pressure with $E_\mathrm{th}/E_k \gtrsim 1$, even in tests with
high-amplitude turbulence perturbation.}
The cores are often not spherical, so we define an effective radius 
of a core with volume $V_\mathrm{core}$ as 
$r_\mathrm{core} = (\frac{3}{4\pi}V_\mathrm{core})^{1/3}$.
For the core properties such as
mass, radius, and density, we use the notation $M_\mathrm{core}$,
$r_\mathrm{core}$, and $\rho_\mathrm{core}$ for cores defined by gravitational
potential alone, and $M_\mathrm{coreb}$, $r_\mathrm{coreb}$, and $\rho_\mathrm{coreb}$
for bound cores with $E_{\mathrm{th}}+E_G<0$. 

To compare the physical properties of cores at different stages of evolution,
two sets of cores are identified using different strategies:
(1) In each simulation for a snapshot at time $t_1$ when the first
core collapses, the cores are identified around each local minimum of the
gravitational potential, similar to \citetalias{GO2011}. 
(2) For each sink particle that forms in the simulation, a corresponding core
is identified at $t_\mathrm{coll}$, the time just before sink particle
formation.
The cores in (2) are by definition gravitationally unstable, and represent the
collapsed stage when the asymptotic LP density profile has developed, just
before a protostar would form.
These cores identified by method (1) represent the structures in the dense regions of
GMCs at an earlier stage of their individual evolution than the cores in method
(2). Not all of the method (1) cores are bound,
nor are they all guaranteed to eventually collapse. For brevity, we use the
terms $t_1$-core and $t_\mathrm{coll}$-core respectively hereafter, for cores
found via the different methods.

\subsection{Model Parameters\label{section:model_parameters}}
To investigate any possible dependence on purely
numerical parameters, we carried out a series of simulations with different
box size and resolution for each Mach number. For each model with given
box size and resolution, we ran between 2 and 8 simulations with different random
initial velocity field for the turbulence perturbations, so that statistics
could be obtained for more than $\sim 100$ cores. 
Because of the planer converging flow geometry, we chose our simulation
box with
$L_x=L_y > L_z$ to focus on the thin post-shock layer. For all simulations,
$L_z=0.625L_J$, and we have tested that using larger $L_z$ does not change the
results significantly. The number of zones in each direction is set such that
each cell is cubic: 
$L_x/N_x=L_y/N_y=L_z/N_z$. The simulations were run until time $t_{\mathrm{lim}}$
when the core formation rate starts to drop significantly because the material
in the post-shock layer is mostly used up by sink particle accretion. As shown in
\citetalias{GO2011} and also \S \ref{section:evolution} in this paper, 
the cores form faster in
higher Mach number simulations, and therefore we chose shorter $t_{\mathrm{lim}}$
for higher Mach number.

Table \ref{table:model_parameters} summarises the model parameters for our study.
We mark in boldface the 
simulation set at each Mach number with the highest resolution and biggest box size,
which will be used for further analysis 
in \S \ref{section:evolution} and \S \ref{section:core} if not
specified otherwise.  Models with smaller box size and resolution are mainly used to
confirm convergence, as discussed in \S \ref{section:convergence}.

As discussed in Section \ref{section:units}, the models of Table
\ref{table:model_parameters} have turbulence levels that depend on Mach number
following Equation (\ref{eq:dv_1D}). In addition, we have conducted two
simulations analogous to model M08L2N256, but with turbulent perturbations an
order of magnitude higher.

\begin{table*}[htbp]
    \centering
    \caption{Simulation Set Model Parameters \label{table:model_parameters}}
    \begin{tabular}{l l c l l l}
        \tableline
        \tableline
         \multicolumn{1}{l}{\tablenotemark{a}Name}
        & ${\cal M}$ &number of simulations &box size ($L_0$) &resolution &
         $t_{\mathrm{lim}}$ ($t_0$)\\
        \tableline
        \textbf{M02L6N256} &\textbf{2} &\textbf{3}  &$\mathbf{6\times6\times0.625}$
        &$\mathbf{1536\times1536\times160}$ &\textbf{0.5}\\
        M02L4N256 &2 &6  &$4\times4\times0.625$  &$1024\times1024\times160$ &0.5\\
        M02L4N128 &2 &4  &$4\times4\times0.625$  &$512\times512\times80$ &0.5\\
        M02L4N64 &2 &8  &$4\times4\times0.625$  &$256\times256\times40$ &0.5\\
        \tableline
        \textbf{M04L6N256} &\textbf{4} &\textbf{2}  &$\mathbf{6\times6\times0.625}$
        &$\mathbf{1536\times1536\times160}$ &\textbf{0.4}\\
        M04L4N256 &4 &3  &$4\times4\times0.625$  &$1024\times1024\times160$ &0.4\\
        M04L4N128 &4 &3  &$4\times4\times0.625$  &$512\times512\times80$ &0.4\\
        \tableline
        \textbf{M08L2N512} &\textbf{8} &\textbf{8}  &$\mathbf{2\times2\times0.625}$
        &$\mathbf{1024\times1024\times320}$ &\textbf{0.35}\\
        M08L2N256 &8 &6  &$2\times2\times0.625$  &$512\times512\times160$ &0.35\\
        M08L2N128 &8 &5  &$2\times2\times0.625$  &$256\times256\times80$ &0.35\\
        M08L4N256 &8 &2  &$4\times4\times0.625$  &$1024\times1024\times320$ &0.35\\
        \tableline
        \textbf{M16L2N512} &\textbf{16} &\textbf{4}  &$\mathbf{2\times2\times0.625}$
        &$\mathbf{1024\times1024\times320}$ &\textbf{0.2}\\
        M16L1.5N512 &16 &6  &$1.5\times1.5\times0.625$  &$768\times768\times320$ &0.2\\
        M16L2N256 &16 &3  &$2\times2\times0.625$  &$512\times512\times160$ &0.2\\
        \tableline
    \end{tabular}
    \tablenotetext{1}{The naming convention is based on the Mach number,
    the horizontal box size in units of $L_0$, and the number of zones in $L_0$.}
\end{table*}

\section{Results}\label{section:results}

\subsection{Overall evolution\label{section:evolution}}
\subsubsection{Development of Cores and Filaments\label{section:filaments}}

\begin{figure*}[htbp]
\centering
\includegraphics[width=\linewidth]{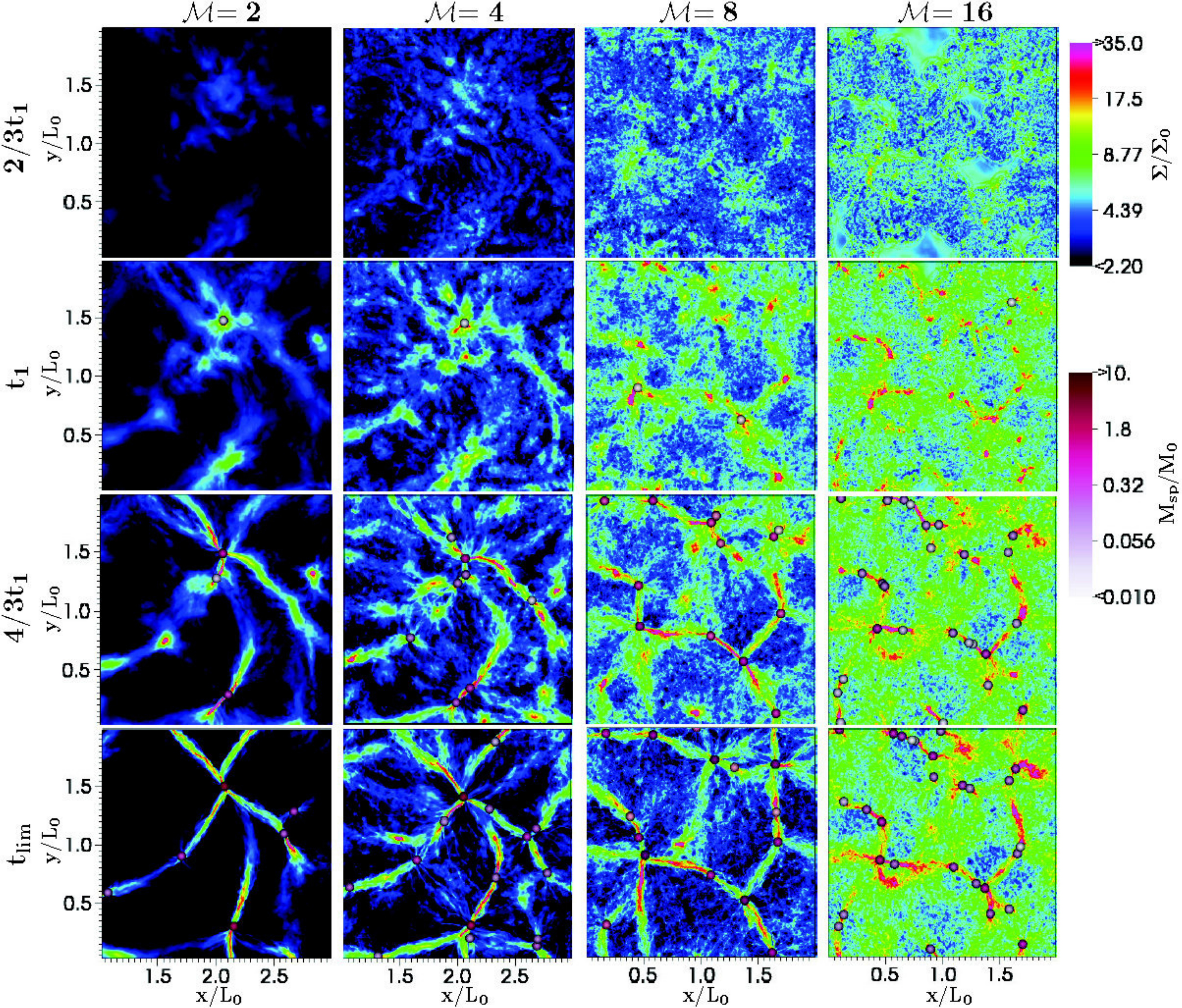}
\caption{Evolution of surface density projected along the direction of
    converging flow ($\hat{z}$).  Logarithmic color scale shows
    $\Sigma/\Sigma_0$ (see Equation (\ref{eq:sigma}) and Table
    \ref{table:units}) for Mach number 2, 4, 8,
16 at times $2/3t_1$, $t_1$,
$4/3t_1$, $t_\mathrm{lim}$. The x- and y- axes are in units of
$L_0=L_J$ (see Equation (\ref{eq:L_J}) and Table \ref{table:units}). 
For ${\cal M}=2,4$, we show only a
part of the simulation domain (a square region with a size of $2L_0$) for direct
comparison with ${\cal M}=8,16$ simulations. 
${\cal M}=2,4$ and ${\cal M}=8,16$ simulations have the
same seeds for turbulent perturbations, and this leads to similar large-scale
structure. The sink particles are over-plotted as
spheres with color scales indicating their masses (note that the plotted size
exceeds the size of the sink control volume).}
\label{fig:evolution}
\end{figure*}

First, we provide an overview of the dynamical evolution in our models.
Figure \ref{fig:evolution} shows the surface density evolution of different
Mach numbers at $(2/3)t_1$, $t_1$,
$(4/3)t_1$, $t_\mathrm{lim}$, where $t_1$ is the time
for the first core to collapse in each simulation, corresponding to $0.28t_0,
0.23t_0, 0.21t_0, 0.13t_0$ for ${\cal M}=2, 4, 8, 16$. The dependence of
$t_1$ on  ${\cal M}$ is discussed below.

The notable density structures seen in all models are filaments and cores. Also
shown in Figure \ref{fig:evolution} are the locations of sink particles.
Because gaseous filaments are unstable to longitudinal self-gravitating
fragmentation \citep{Nagasawa1987}, the suggestion has been that there is a
two-stage process for core formation, with filaments first forming, and then
fragmenting gravitationally \citep[e.g.,][]{Andre2014}. However, we find 
a subtly different progression.
Figure \ref{fig:evolution} makes clear that cores and filaments develop
simultaneously, rather than filaments forming first, followed by fragmentation
into cores. Snapshots at closely-space time intervals show that the central
densities of proto-cores and filaments grow together,
until some of the gravitationally unstable cores
collapse and form sink particles. As a result, the sink particles are not
randomly distributed in the post-shock region, but are dotted along the long
thin filamentary over-dense structures. Because our simulations do not include
feedback to limit accretion, sink particles continually accrete surrounding material,
and after some time may merge with one another. 

The similarity in the density structure patterns of ${\cal M}=2,4$ or ${\cal
M}=8,16$ arises because they have the same seeds for turbulence perturbations.
Because of the
planar converging flow geometry, the $\hat{x}$ and $\hat{y}$ component of 
the large scale velocity are not
changed by the shock, and these velocities seed the density fluctuation patterns
that grow to create filaments and cores. Thus, the structure provided by
turbulence is crucial to initiation of the self-gravitating cores that collapse
to make sinks. However, it is
not true that the cores are formed simply from the growth of initial
perturbations. Filaments and clumps can merge and fragment before they finally
collapse into sink particles. 

For the present simulations, the initial
turbulent motions are generally subsonic. Although the structures that grow
are seeded by the turbulent compression, self-gravity is crucial to their
development\footnote{Even for the models with larger-amplitude turbulence,
gravity is important for creating core-forming filaments.}.
This is true for both filaments and cores. We note, however, that
on larger scales for which flow velocity would be supersonic, filamentary
structures can form from turbulent compression alone. This sort of filament,
for which gravity is not important, is distinct from the filaments formed in
our simulations\footnote{\citet{Andre2010} discusses observations of filaments in
both non-self-gravitating and self-gravitating clouds.}.

\begin{figure*}[htbp]
\centering
\includegraphics[width=0.7\linewidth]{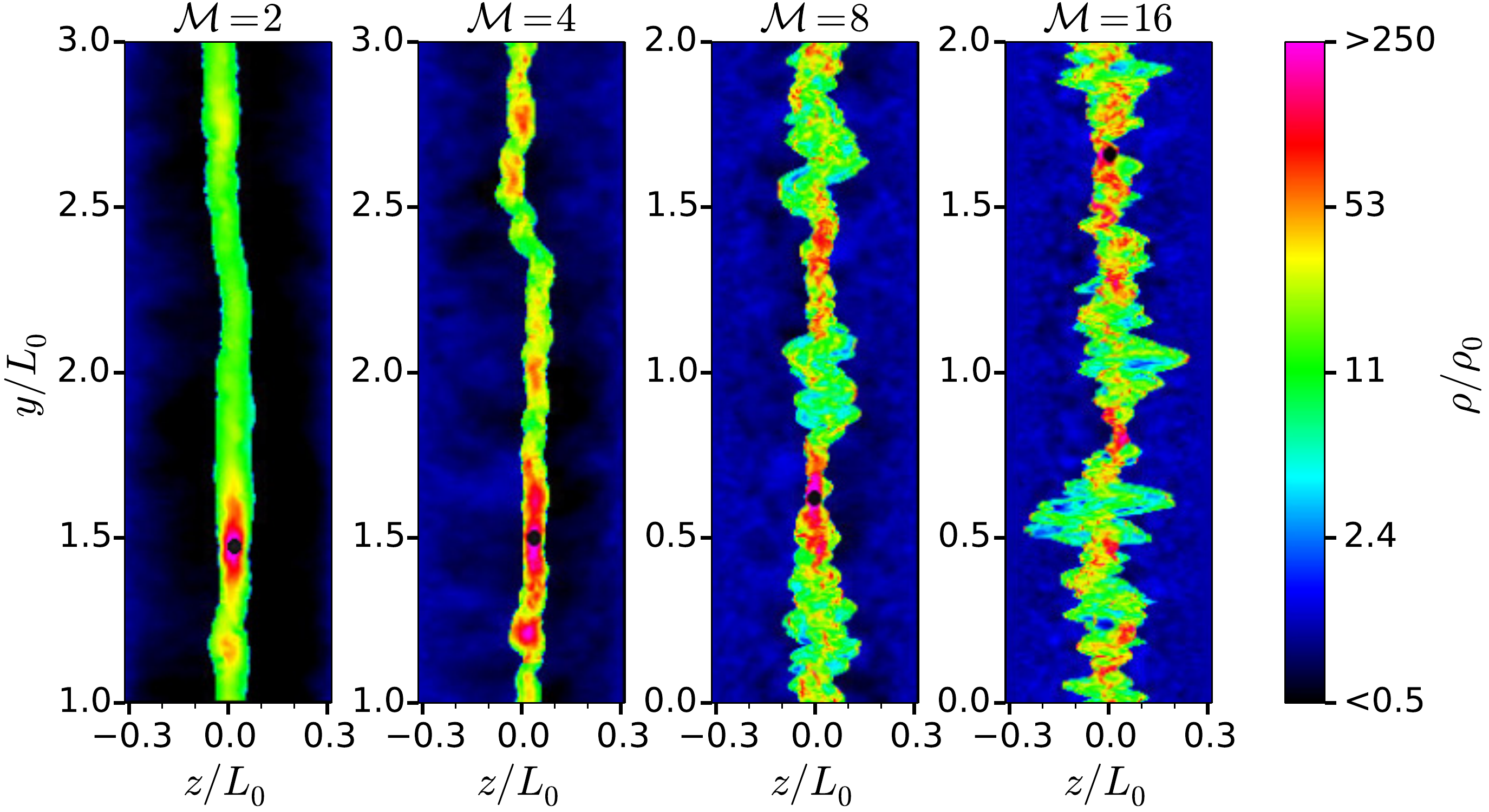}
\caption{Slice of density along the x-axis at the time and position of the
first sink particle. The sink particle in each panel is marked with a black dot.
Notice that the
higher ${\cal M}$ models have strongly dispersed post-shock layers (along the
inflow direction $\hat{z}$), as a consequence of hydrodynamic instability of the
shock-bounded layer.}
\label{fig:evolution_slice}
\end{figure*}

Figure \ref{fig:evolution_slice} shows an $x=const$ slice through the
density maximum at the time when collapse leads to the first sink particle. 
Evidently, the post-shock layer seen in cross section can be highly nonuniform,
especially for ${\cal M}=8$ and $16$. This is due to nonlinear thin-shell 
instabilities of the post-shock layer \citep{Vishniac1994}
that are stronger at higher Mach number.
As a result of these instabilities, the gas is vertically dispersed and the
density in the post-shock region is lower than $\rho_0{\cal M}^2$, the value
that would apply for a simple isothermal shock.
The thickness of the post-shock layer is correspondingly higher than the
value that would apply in the absence of instabilities.
In fact, at later times, the median density (and pressure) of the 
post-shock region is nearly independent of Mach number. 
This leads to a very shallow dependence of core mass on Mach
number, which is discussed further in \S \ref{section:core}.
We note that magnetic fields reduce instabilities of the post-shock
region \citep{CO2014}, so that the post-shock density and magnetic pressure
increase with inflow Mach number.

\begin{figure*}[htbp]
\centering
\includegraphics[width=0.85\linewidth]{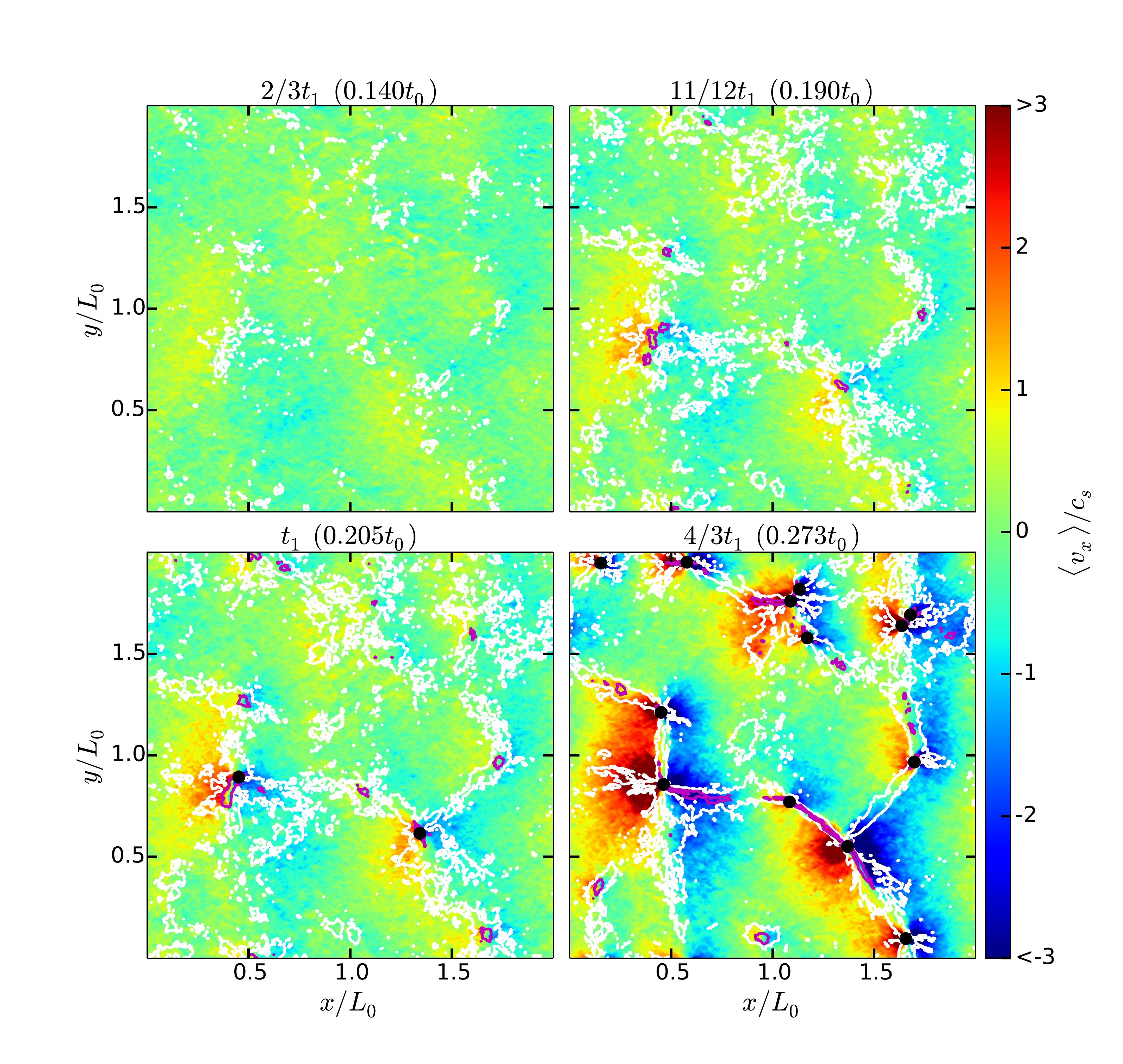}
\caption{Color scale shows maps of $v_x$ averaged over the z-axis:
    $\langle v_x \rangle =\int v_x \rho \di z / \int \rho \di z$, for one
    ${\cal M}=8$ simulation at $t/t_1 =2/3,11/12,1,4/3$. 
    The white and magenta curves are contours of surface density
    $\Sigma/\Sigma_0=8$ and $24$. Sink particles are marked as black dots. 
Gas converges toward the ridges of forming filaments, and then accretes at
free-fall onto sink particles.}
\label{fig:evolution_vx}
\end{figure*}

The map of $v_x$ at different times is shown in Figure
\ref{fig:evolution_vx}. At the early core-building stage, 
the velocities are sub-sonic, converging towards the ridge of growing
filaments \citep[see also \citetalias{GO2011}, ][]{CO2014}, due to the gravity
from cores and filaments. After a core
collapses to reach an LP profile, a sink particle is created and
the surrounding material accretes onto the sink
particle at the free-fall velocity. 

\subsubsection{Collapse of the First Cores: the Non-linear Time and
Mass\label{section:non-linear}}
Treating the post-shock layer as a slab with half thickness $H$ and surface density
$\Sigma=2c_s {\cal M} \rho_0 t$ that grows in time, for linear-amplitude 
in-plane perturbations with $kH \ll 1$,
\citetalias{GO2011} showed the wavelength of the unstable mode that would have the greatest
exponential amplification at a given time is 
\begin{equation}\label{eq:lambda_m}
    \frac{\lambda_m}{L_0} = \left( \frac{2\sqrt{3}}{\Gamma_\mathrm{max}} \right)^{1/2}
               \frac{1}{{\cal M}^{1/2}},
\end{equation}
where $\Gamma_\mathrm{max} = \ln \delta \Sigma_\mathrm{max}/\delta
\Sigma_\mathrm{init}$. The time at which an amplification
$\Gamma_\mathrm{max}$ is reached is
\begin{equation}\label{eq:t_m}
    \frac{t_m}{t_0} = \left( \frac{2\Gamma_\mathrm{max}}{\sqrt{3}\pi^2} \right)^{1/2}
         \frac{1}{{\cal M}^{1/2}}.
\end{equation}
A characteristic mass associated with this mode is
\begin{equation}\label{eq:M_m}
    \frac{M_m}{M_0} = \frac{\left( \frac{\lambda_m}{2} \right)^2 \Sigma(t_m)}{M_0}
        = \left( \frac{2\sqrt{3}}{\Gamma_\mathrm{max}\pi^2} \right)^{1/2}
           \frac{1}{{\cal M}^{1/2}}.
\end{equation}

\begin{figure*}[htbp]
    \plottwo{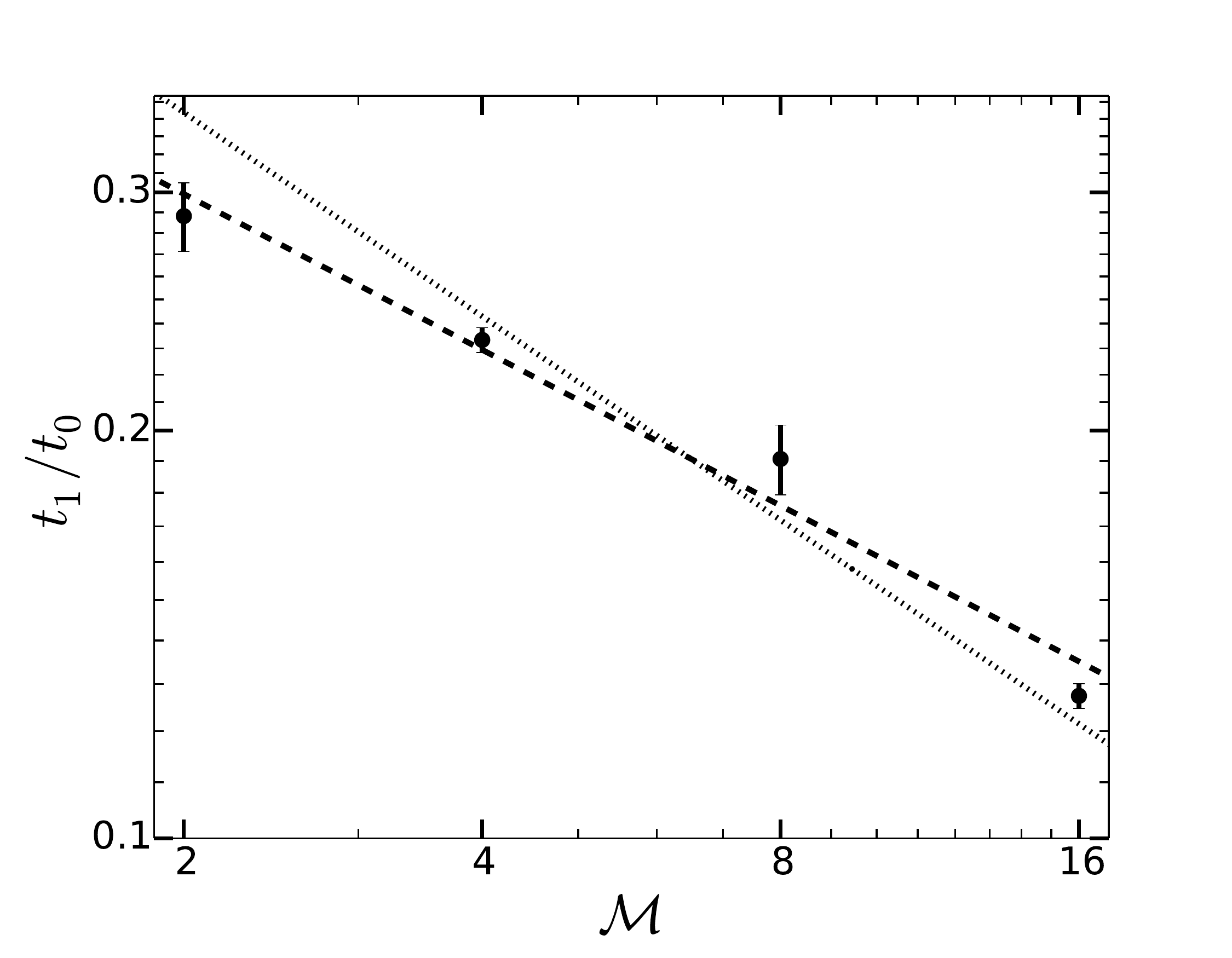}{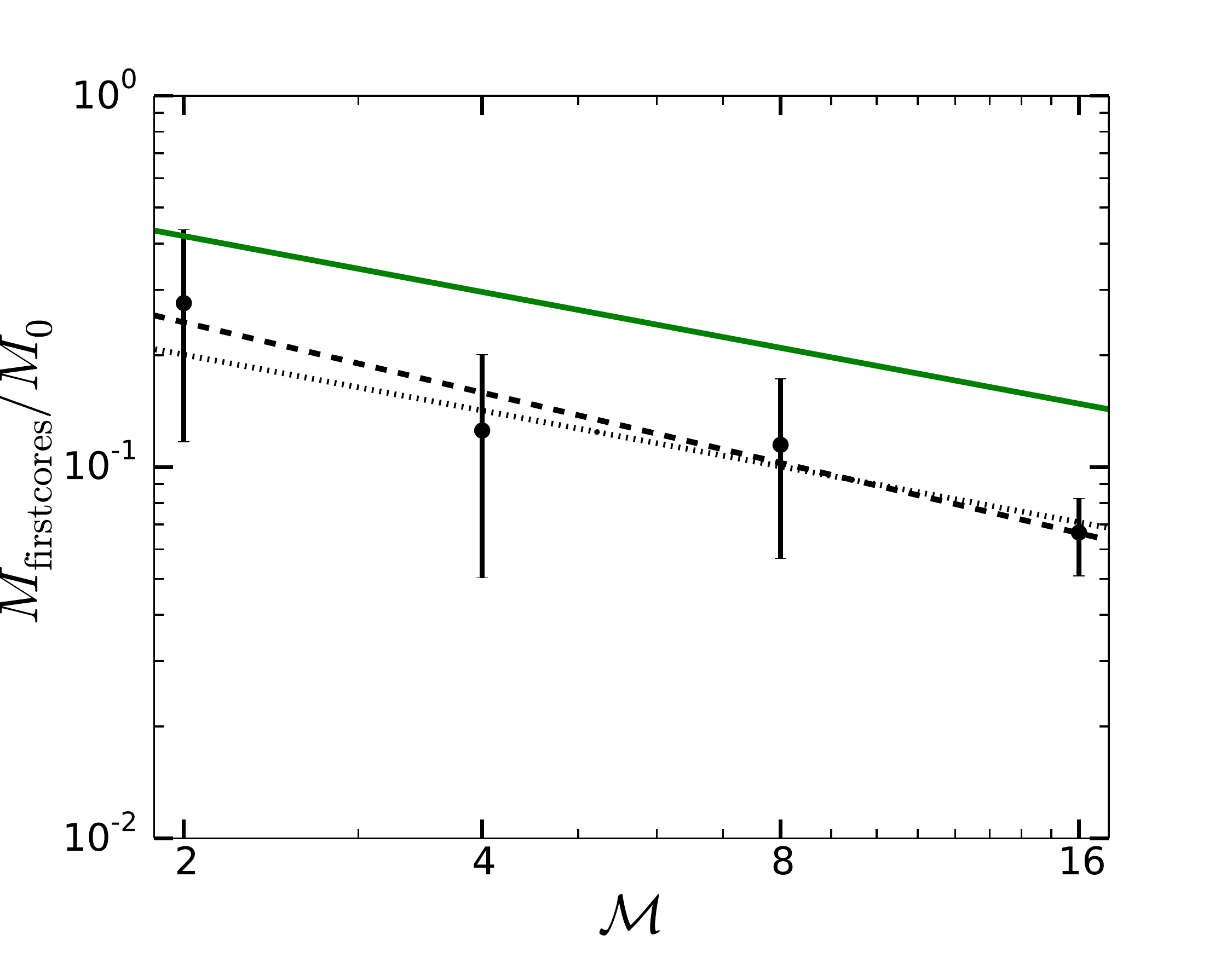}
    \caption{Time of first collapse $t_1$ (left) and average mass 
      $M_\mathrm{first cores}$ (right) of the first cores
    verses Mach number ${\cal M}$. The dots
    are the average values for all the included models (see text), with 
    error bars showing the standard deviations. The dashed
    line is the log linear fit, and the dotted line is the fit with a
    fixed slope of -1/2. The fitting gives $t_1\propto {\cal M}^{-0.38}$
    and $\langle M_\mathrm{first cores} \rangle \propto {\cal M}^{-0.63}$,
    similar to $t_m, M_m\propto
    {\cal M}^{-0.5}$ in equation (\ref{eq:t_m}) and (\ref{eq:M_m}). If using a fixed
    slope of -1/2, the fitting gives $\Gamma_\mathrm{max}=1.3$ for $t_1$ in equation 
    (\ref{eq:t_m}) and $\Gamma_\mathrm{max}=2.4$ for
    $\langle M_\mathrm{first cores} \rangle$ in equation (\ref{eq:M_m}).
    The green solid line in the right panel plots
    $M_m$ in equation (\ref{eq:M_m}) with $\Gamma_\mathrm{max}=1$.
\label{fig:t_m}}
\end{figure*}

We fitted the time for the first core to collapse $t_1$ and the
average mass of the first five cores in each simulation
$\langle M_\mathrm{first cores} \rangle$ as a function of Mach number. 
In addition to the primary models (boldface in Table
\ref{table:model_parameters}), we also included M02L4N1024, M04L4N1024, and
M16L1.5N768 models, which have the same resolution and have been
demonstrated to have sufficient box-size (see \S \ref{section:convergence}).
The result is shown in Figure \ref{fig:t_m}. Our fit gives
$t_1/t_0 = 0.39 {\cal M}^{-0.38}$ and 
$\langle M_\mathrm{first cores} \rangle /M_0 = 0.38 {\cal M}^{-0.63}$.
The dependence of $t_1$ and $\langle M_\mathrm{first cores} \rangle$
on Mach number is similar to the dependence $t_m, M_m\propto {\cal M}^{-0.5}$
of Equations (\ref{eq:t_m}) and (\ref{eq:M_m}), and $\Gamma_\mathrm{max}=1-2$
would give coefficients in Equations (\ref{eq:t_m}) and (\ref{eq:M_m})
comparable to our fits.

However, the large dispersion in $\langle M_\mathrm{first cores} \rangle$ (see
Figure \ref{fig:t_m})
suggests a process more complex than local growth of cores with in-plane waves
$l_x \sim l_y \sim \lambda_m$. In particular, filament development 
with $\l_x$ and $\l_y$ unequal clearly plays a role in core formation. 
We note that using $\Gamma_\mathrm{max}=1$, Equation \ref{eq:lambda_m} gives
$\lambda_m/L_0 = 1.86/{\cal M}^{1/2}$, which we find is roughly equal to the
typical filament separation 
seen in Figure \ref{fig:evolution}, and also consistent with what was found in
\citet{Loo2014}. Thus, both the time to first collapse and the prominence of
filaments implies that non-linear instability of asymmetric 
self-gravitating in-plane modes is important to core formation.

Finally, we note that the mass per unit length associated with the most
amplified mode is
$\lambda_m\Sigma(t_m)=4c_s^2/G$, which is equal to twice the critical mass per
unit length for an isothermal filament \citep{Ostriker1964}, or
$33M_\sun\mathrm{pc}^{-1}$ for $T=10\mathrm{K}$.\footnote{For a self-gravitating
isothermal sheet with fixed surface density $\Sigma$ and scale height $c_s^2/(\pi
G\Sigma)$, the mass per unit length of the fastest growing mode is also
$\approx 4c_s^2/G$.} 
Of course, this mass is not
all available at time $t_m$, because only a fraction of this has been gathered
into the filament. We find that typical values of the mass per unit length in
filaments at time $t_1$ is $\sim (1-3)c_s^2/G$.

\subsubsection{Evolution of Cores}
\begin{figure}[htbp]
    \includegraphics[width=\linewidth]{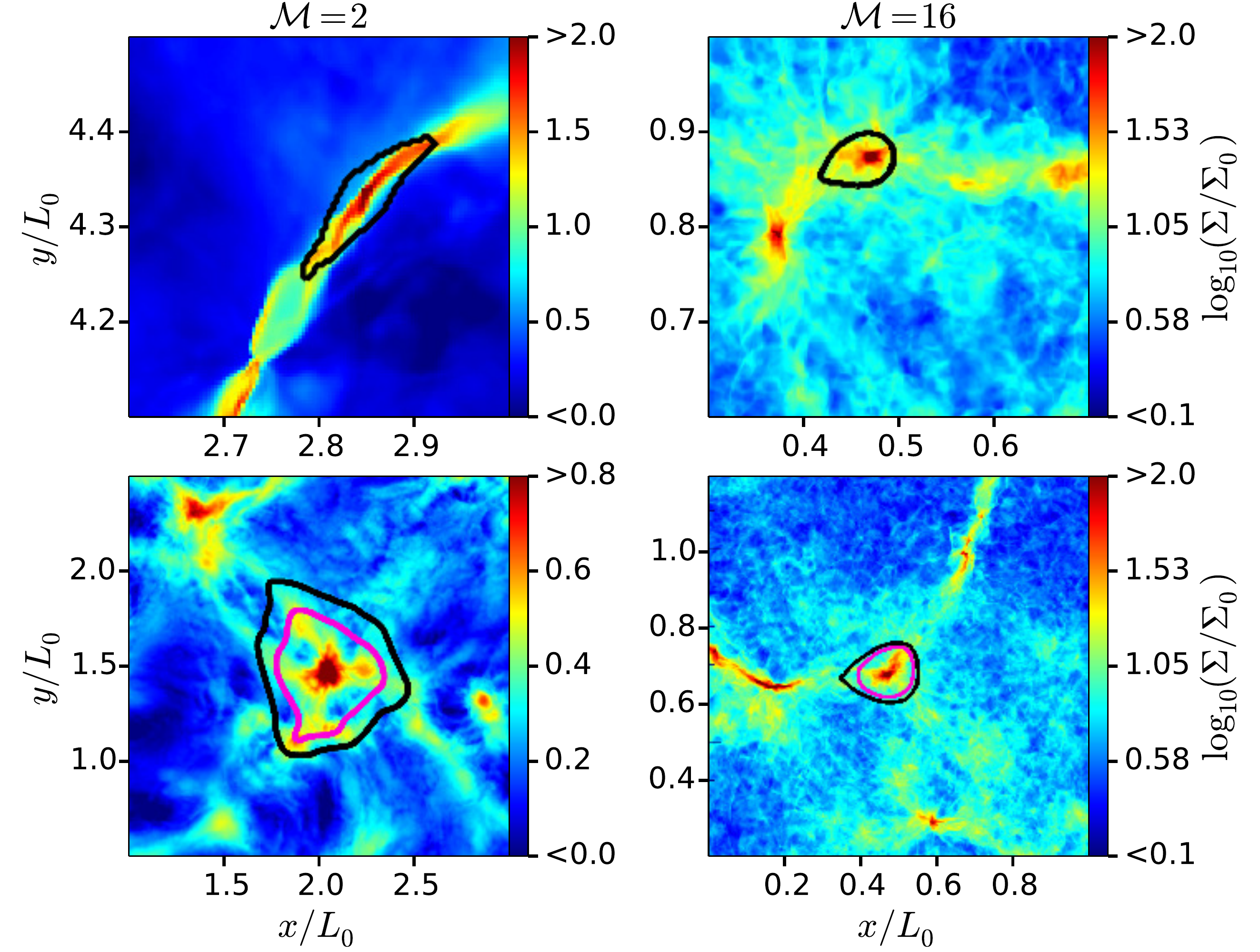}
    \caption{Illustration of individual $t_\mathrm{coll}$-cores, 
    for ${\cal M}=2$ (left panels) and ${\cal M}=16$ (right panels) simulations.
    The upper and lower panels show typical-mass and high-mass cores
    respectively, which are marked as red dots and star symbols in Figure
    \ref{fig:mcore_ti}.
    The color scheme shows surface density $\log_{10}(\Sigma/\Sigma_0)$.
    The black and magenta curves draw the boundaries projected along 
    the z-axis of cores defined by gravitational potential alone, and
bound core defined by $E_{th}+E_G<0$ (not plotted in the upper panels for
clarity).}
    \label{fig:corefind_tcoll}
\end{figure}

\begin{figure}[htbp]
    \includegraphics[width=\linewidth]{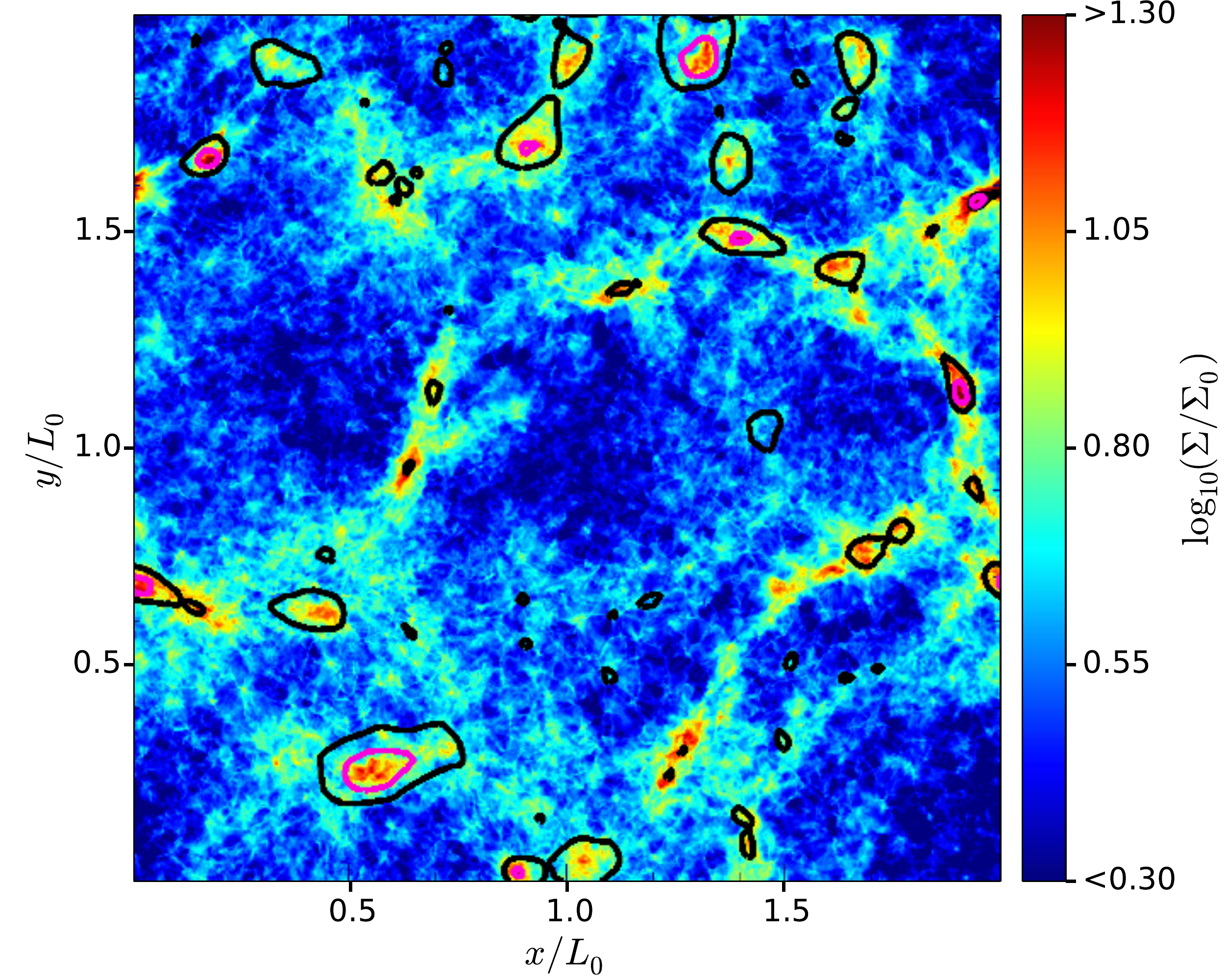}
    \caption{Illustration of $t_1$-cores for one ${\cal M}=8$ simulation.
    Black and magenta curves are as in Figure \ref{fig:corefind_tcoll}.}
    \label{fig:corefind_t0}
\end{figure}

Examples of core finding 
for individual $t_\mathrm{coll}$-cores and an overall map of all $t_1$-cores in
one simulation are shown 
in Figures \ref{fig:corefind_tcoll} and \ref{fig:corefind_t0}.  
We can see clearly that cores are associated with filaments.
A typical mass core is often found embedded in dense filaments. Although high
mass cores may form in relative isolation early on, these locations often
become junctions of filaments later on. These structures of cores lying in
filaments or at the junctions are qualitatively very similar to the observation
of star-forming regions with {\sl Herschel} \citep{Andre2014}, as well as
previous work \citep[e.g.,][]{JB1999, Hartmann2002}.

\begin{figure*}[htbp]
\centering
\includegraphics[width=0.9\linewidth]{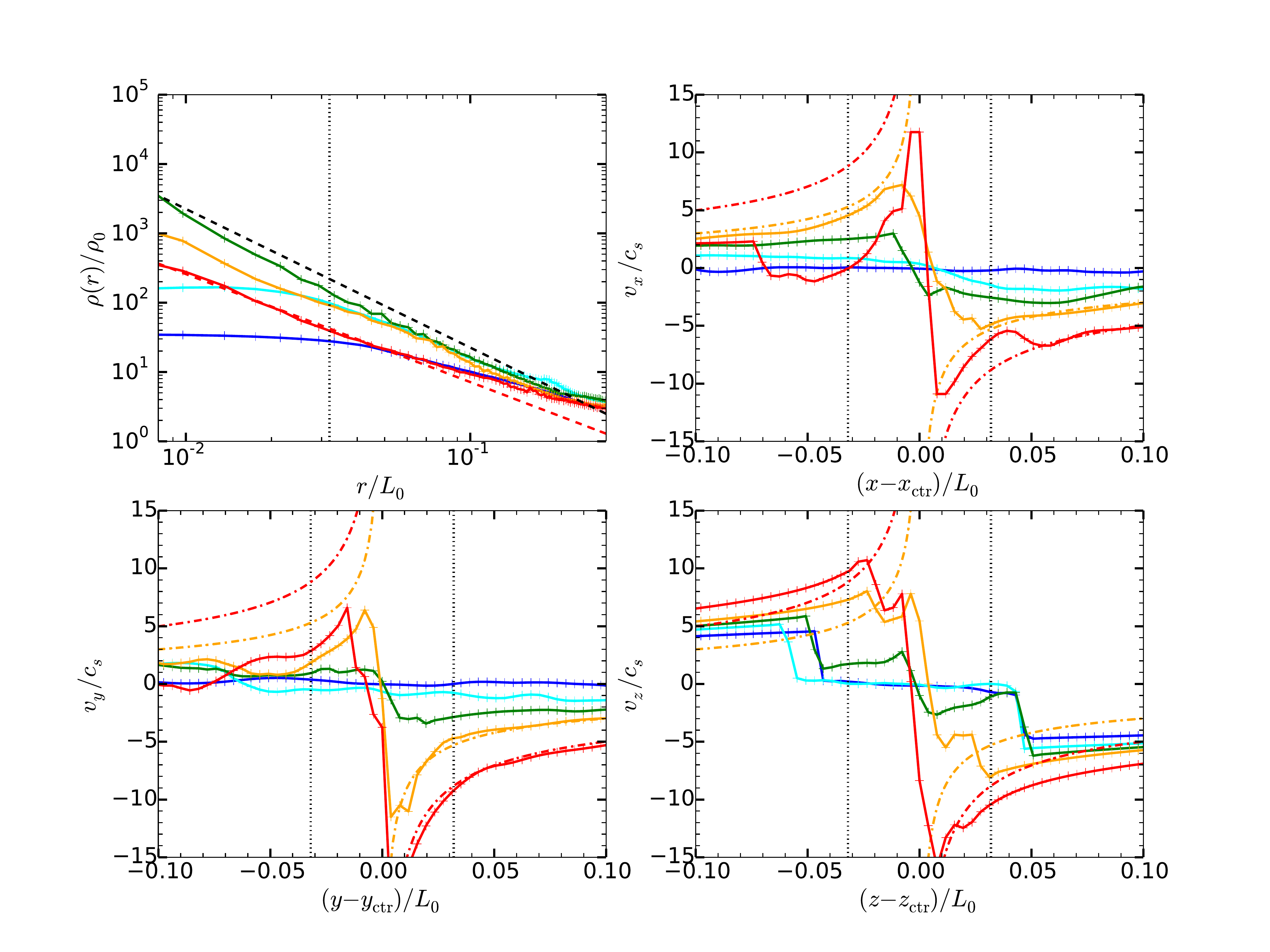}
\caption{Density and velocity evolution of a typical mass core in one 
    ${\cal M}=2$ simulation 
    (marked as a red sphere in the upper left panel of Figure
    \ref{fig:mcore_ti} and shown in the upper left of Figure
    \ref{fig:corefind_tcoll}). 
The blue, cyan, green, orange and red lines show the core
profiles at $t/t_0=$0.260, 0.343, 0.373 ($t_\mathrm{coll}$), 
0.403, and 0.467. The crosses mark on each profile 
the grid centers in the simulations.
This sink particle will merge with another sink at
$t_\mathrm{merge}/t_0=0.483$.
Upper left: angle-averaged density along $r$. The black dashed line
shows the LP profile (Equation \ref{eq:rho_LP}).
The red dashed line is a log-linear fit of the red density profile 
for $2\Delta x < r < 2r_\mathrm{core}$, 
which gives  $\rho(r)\propto r^{-1.56}$.
Upper right, lower left and right: $v_x$, $v_y$ and $v_z$ along the $\hat{x}$,
$\hat{y}$ and $\hat{z}$ axis. 
The orange and red dash-dotted lines show the free-fall velocity $v_{ff} =
\sqrt{2GM_\mathrm{sp}/r}$, where $M_\mathrm{sp}$ is the mass of the sink
particle at the corresponding time of the orange and red velocity profiles. 
The dotted vertical lines
denote the value of $r_\mathrm{core} = [3V_\mathrm{core}/(4\pi)]^{1/3}$ at $t_\mathrm{coll}$.}
\label{fig:core_propertiesM02}
\end{figure*}

\begin{figure*}[htbp]
\centering
\includegraphics[width=0.9\linewidth]{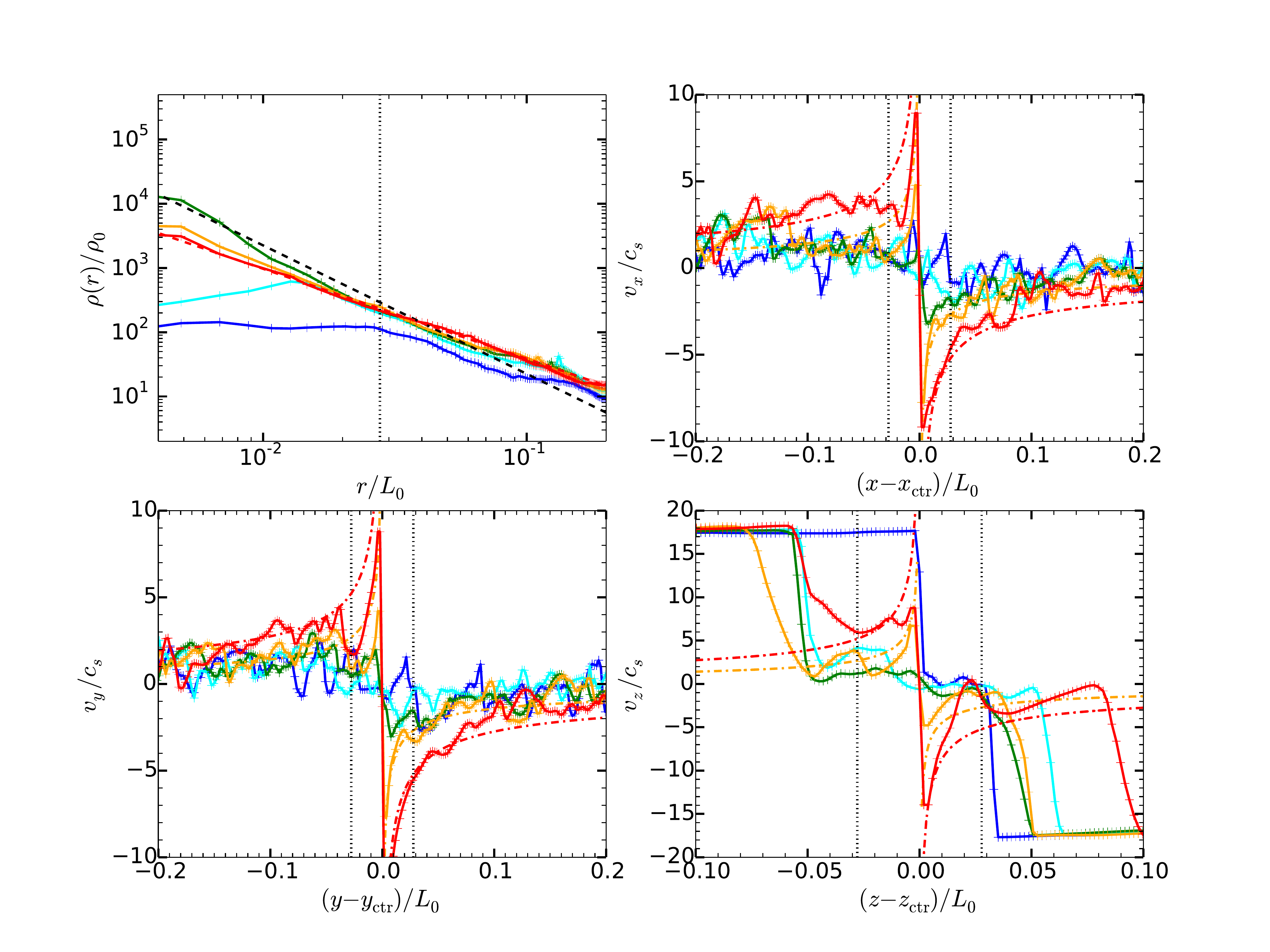}
\caption{Same as Figure \ref{fig:core_propertiesM02} for a typical core in one
${\cal M}=16$ simulation run 
(marked as a red sphere in the lower right panel of Figure \ref{fig:mcore_ti}
and shown in the upper right of Figure \ref{fig:corefind_tcoll}) at
$t/t_0=$0.100, 0.134, 0.146 ($t_\mathrm{coll}$), 0.151 and 0.165.
This sink particle will merge with another sink at
$t_\mathrm{merge}/t_0=0.171$.
Upper left: The red dashed line gives a fit of $\rho(r)\propto r^{-1.40}$ for
the density profile during the infall stage. 
The velocities are more turbulent than in the ${\cal M}=2$ model, but
still mostly subsonic within the core before it collapses.}
\label{fig:core_propertiesM16}
\end{figure*}

\citetalias{GO2009} classified the core development into four different stages:
core building, core collapse, envelope infall, and late accretion. These stages can
also be seen in our simulations, as illustrated in Figure
\ref{fig:core_propertiesM02} and \ref{fig:core_propertiesM16} 
of the density (angle-averaged) and velocity profiles of typical cores.
Initially at the core building stage, the core evolves in quasi hydrostatic
equilibrium, with subsonic internal velocities.
The core mass and density grow slowly as
the gas from surrounding environment flows into the core potential well.
When the core becomes gravitationally unstable, it collapses
in a short timescale, reaching the LP profile.
Then a sink particle is created in the core center, 
and starts accreting the gas from the envelope, causing 
the density to drop as the core collapses inside-out 
and approaches $\rho(r)\propto r^{-1.5}$, a profile close to
the expectation for free-fall \citep{Shu1977}. 

Noticeably, the angle-averaged core density profiles continue to drop smoothly
beyond the effective core radius
$r_\mathrm{core}=[3V_\mathrm{core}/(4\pi)]^{1/3}$ as shown in 
Figure \ref{fig:core_propertiesM02} and \ref{fig:core_propertiesM16}.
We note, however, this is partly due to the method we use to
calculate $\bar{\rho}(r)$, which takes the average density in a spherical shell 
at distance $r$ from the core center, including the low-density
pre-shock gas at $r/L_0 \gtrsim 0.04$ outside of the planer post-shock layer.
For example, the density profile
of the core in Figure \ref{fig:core_propertiesM16} only extends to $\sim 2
r_\mathrm{core}$ along some directions and is much less smooth than the
angle-averaged density profile, as can be seen in the upper right of
Figure \ref{fig:corefind_tcoll}.
The anisotropy of cores (e.g., as in the upper left of Figure \ref{fig:corefind_tcoll})
can also make the angle-averaged density differ from the density profile 
along individual principal axes. 
Nevertheless, cores often extend beyond their effective radius $r_\mathrm{core}$,
because they are formed in denser environments than the
average post-shock density, and instantaneous tidal forces (which define
$r_\mathrm{core}$) do not limit
the material that may ultimately fall into a sink particle.
In fact, we often see that the mass of the sink continues to grow 
after it reaches the mass of the $t_\mathrm{coll}$-core. 
There is no obvious boundary between the envelope infall and late accretion
stages, as discussed in \S \ref{section:accretion}.

In our tests of the case with high-amplitude perturbations,
the overall evolution is very
similar to the low-amplitude cases. Cores and filaments still form in a
similar manner, with cores embedded within filaments.
At smaller scale ($l \ll L_J$), both the initial
velocity perturbation in Equation (\ref{eq:dv}) is smaller and the turbulence 
dissipation timescale $t = l/v \propto l^{1/2}$ is shorter. As a result, 
even when the large-scale velocity perturbation
is super-sonic, the high density regions within filaments where cores form are generally
sub-sonic or trans-sonic, leading to very similar core evolution dominated by
thermal pressure and gravity. Individual cores formed in simulations with
high-amplitude
velocity perturbations show the same internal collapse and infall profiles as
shown in Figures \ref{fig:core_propertiesM02} and \ref{fig:core_propertiesM16}.

\subsection{Convergence of isothermal fragmentation\label{section:convergence}}

In order to understand protostar formation in our simulations, it is
important to test numerical convergence, both to make sure the fragmentation
process is not purely numerical, and the cores in our simulations are well
resolved. In the literature, there has been a debate of whether 
fragmentation in isothermal HD simulations
is determined by purely numerical rather than physical factors. If an
isothermal sphere were to collapse homologously,  the Jeans length within the
sphere would decrease with radius as $\propto R_\mathrm{sphere}^{3/2}$.
This led to the early idea \citep{Hoyle1953} of hierarchical
fragmentation into smaller pieces until pieces are no longer isothermal. 
In reality, however, core collapse is highly nonuniform and leads to
centrally-concentrated structures \citep{BS1968, Larson1969, Penston1969}.
At every radius in the asymptotic LP solution, the Jeans length is comparable
to the radius.
Thus it is not obvious that fragmentation can happen at arbitrarily small scales
even for a simple isothermal equation of state.
\citet{Martel2006} carried out a series of isothermal SPH simulations employing
particle splitting techniques, and argued that the core masses they found
are determined by numerical resolution and the density threshold of sink particles. 
However, the
highest resolution they used only marginally resolves the Jeans mass at the density
threshold for their sink particles, and they did not test with a higher
resolution or density threshold.
\citet{Krumholz2014} further suggested that isothermal evolution with
self-gravity cannot lead to a characteristic distribution of fragment mases, and
that stellar masses must depend crucially on non-isothermal effects.
However, \citet{IM1997} found that filaments (even with supercritical mass per
unit length $>2c_s^2/G$) fragment longitudinally (rather than collapsing to a
spindle) provided that the initial density perturbations are sufficiently large. In real
clouds, as cores and filaments begin their growth together, filaments may not
become supercritical until after cores have become nonlinear.

\begin{figure*}[htbp]
\centering
\includegraphics[width=0.9\linewidth]{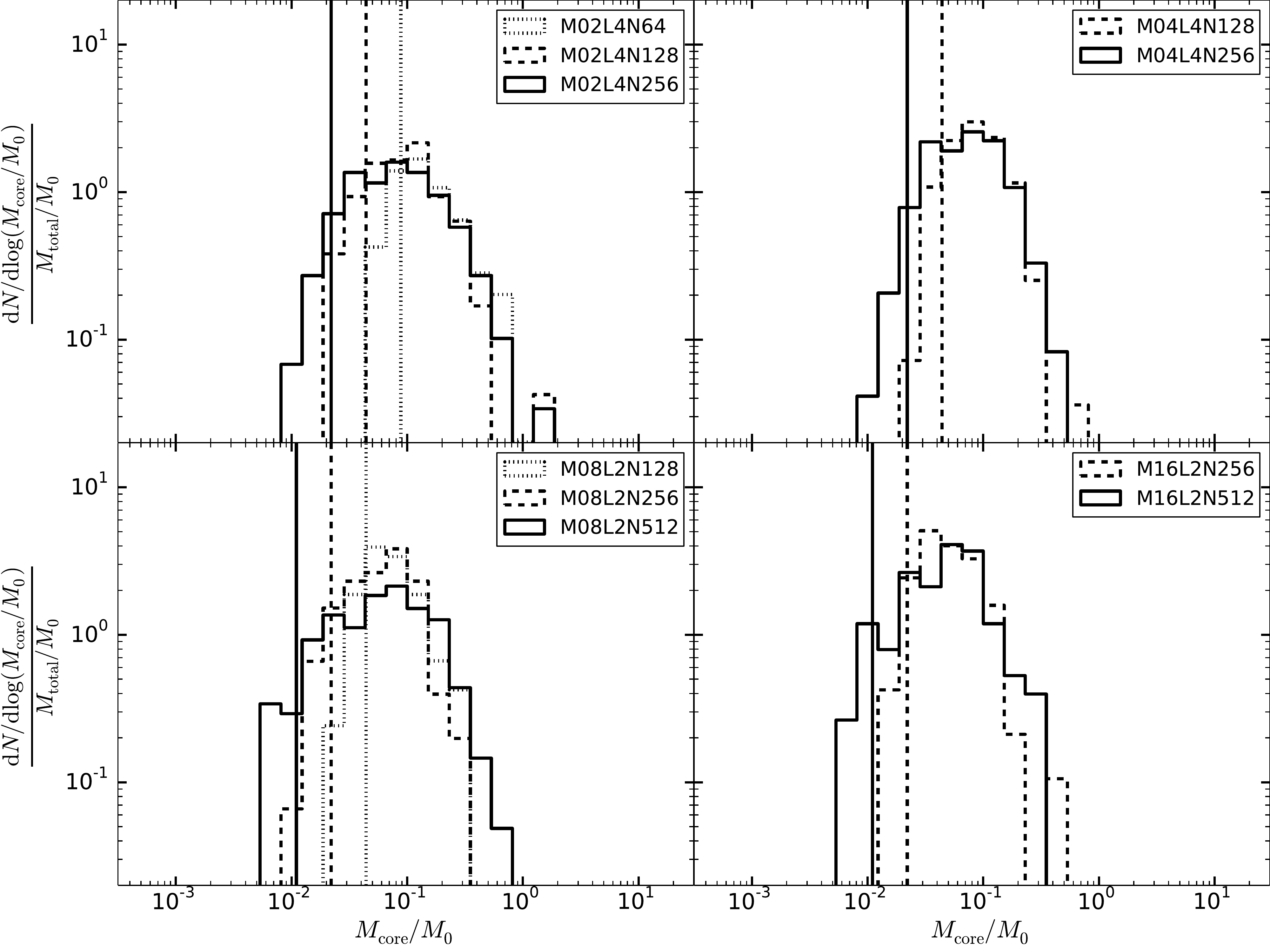}
\caption{The core mass function (CMF) for $t_\mathrm{coll}$-cores 
of models with different resolutions at each Mach number. 
Here the cores are defined by gravitational potential alone.
The y-axis is normalized such that the area of each histogram is
one. The vertical lines are $M_\mathrm{LP}(2\Delta x)$
in equation (\ref{eq:M_LP}), on the left of which the cores are considered not
well resolved. 
There are about 140-300 cores in each model. Each histogram is normalized so
that the total mass equals $M_0$. Although the minimum core mass decreases
with increasing resolution, the peak of the CMF does not.}
\label{fig:convergence_Mcore}
\end{figure*}

\begin{figure*}[htbp]
\centering
\includegraphics[width=0.9\linewidth]{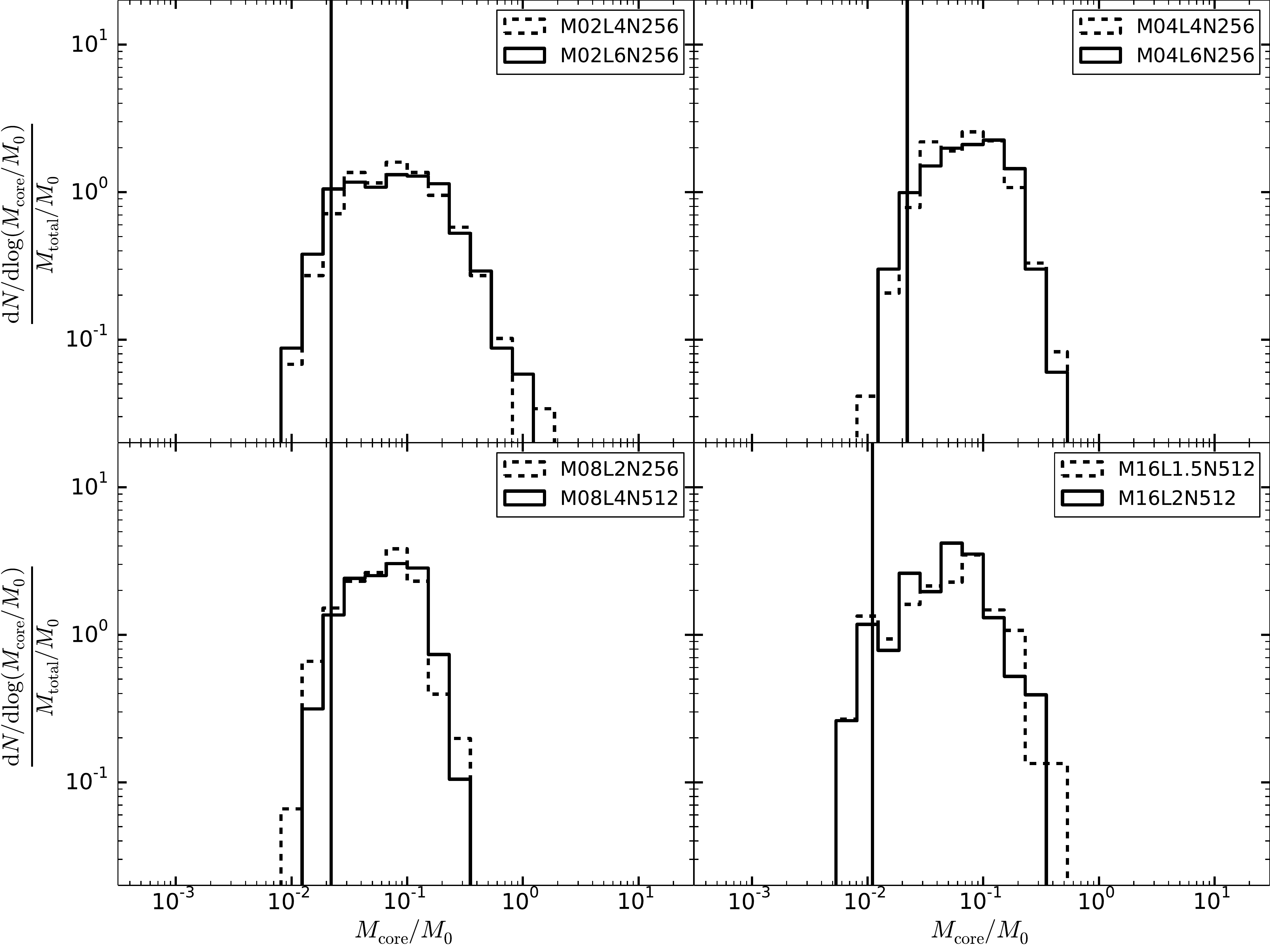}
\caption{Similar to Figure \ref{fig:convergence_Mcore} for models with different
box sizes.}
\label{fig:convergence_box}
\end{figure*}

We find that {\it at the stage of core collapse}, there is in fact
a well defined characteristic mass scale for the ideal isothermal case. 
Figure \ref{fig:convergence_Mcore}
shows the core mass function (CMF) for $t_\mathrm{coll}$-cores
of models with different resolutions for each
Mach number. For the lower resolution models, the distribution of core mass drops
off rapidly at $r_\mathrm{core} \lesssim 2\Delta x$. 
This shows that cores with $\lesssim  12$ zones may not be
well resolved. However, as we increase the resolution, although the CMF
continues to extend towards lower masses,
the peak of the CMF does not vary much (see also Table \ref{table:core_properties}). 
We note that the density threshold for sink particle creation
also increases with resolution as described in Equation (\ref{eq:rho_thr}).
For the highest resolution model of each Mach number, 
the CMF peaks at a mass of cores that are well resolved with
$r_\mathrm{core}/\Delta x = 7-13$.
Although simulations with even higher resolutions could show a modification of
the CMF, especially at the low mass end, our models suggest that the CMF peak, 
or the characteristic mass of the core, would remain the same.

We also investigated the numerical effect of horizontal box size $L$ in our
simulation. The box size limits the longest
wavelength perturbation that can grow in the simulation domain
$\lambda < L$, and also limits the amplitude of initial velocity
perturbations as described in Equation (\ref{eq:dv_1D}).
Figure \ref{fig:convergence_box} shows histograms of $M_\mathrm{core}$ 
of models with different box sizes. The distributions of $M_\mathrm{core}$ 
do not vary appreciably with increasing box size in our models 
(see also Table \ref{table:core_properties}). This shows that
for our simulations with a box sufficiently larger than the dominant non-linear 
wavelength ($L > \lambda_m$, see Equation \ref{eq:lambda_m}), 
the box size does not have a significant effect on core formation.

Although our simulations show a well-defined, converged value for the peak in
the distribution of core masses at the point of singularity formation,
fragmentation at later stages could in principle further alter this
distribution. In particular, infall of rotating envelopes leads to formation
of prestellar accretion disks, which we do not model in the current study.
These accretion disks would be susceptible to fragmentation if external and internal
heating are not properly taken into account. We conclude that it is important
to distinguish between the core stage and subsequent stages in assessing the
outcome of self-gravitating fragmentation in turbulent clouds. Going from the CMF
to IMF physically involves structures at increasingly high temperature, and
simulations must include these non-isothermal effects to model the stages of
disk formation and evolution.

\subsection{Core Properties\label{section:core}}
\subsubsection{Typical Core Mass, Radius, and Density}
We summarise the basic physical properties for
$t_\mathrm{coll}$-cores and $t_1$-cores
of different models in Table \ref{table:core_properties} and Table
\ref{table:core_properties_t0}.  
Cores containing fewer than 27 grid cells are considered
to be poorly resolved (similar to the resolution limit in Figure
\ref{fig:convergence_Mcore}) and are not included in 
Table \ref{table:core_properties} or further analysis.
Table \ref{table:core_properties} again shows the convergence of 
the characteristic mass for
simulations with different resolutions and box
sizes, as discussed in \S \ref{section:convergence}.

\begin{figure*}[htbp]
\centering
\includegraphics[width=0.8\linewidth]{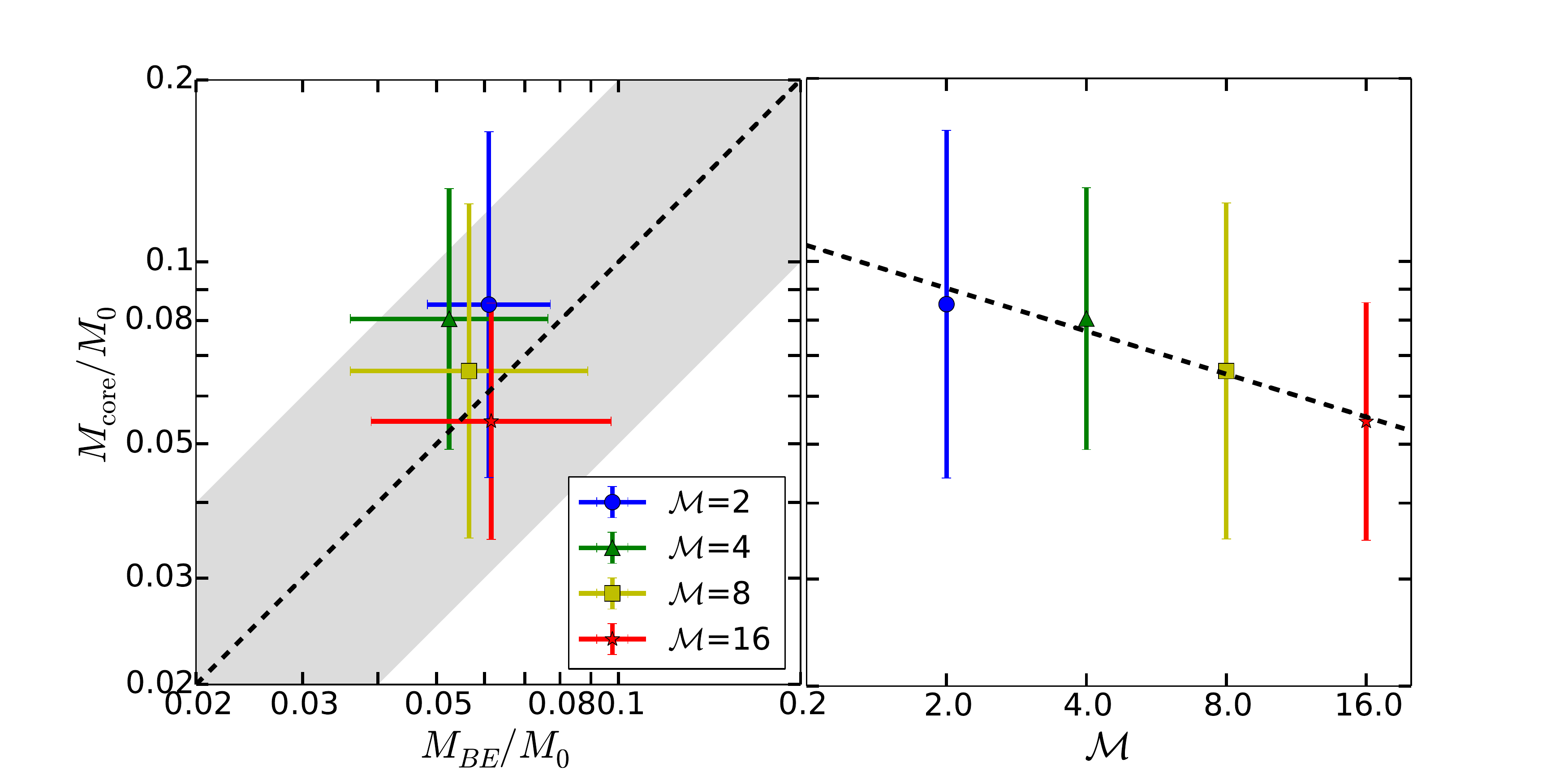}
\caption{Left: Median mass
    of $t_\mathrm{coll}$-cores ($M_\mathrm{core}$) 
    compared to the post shock critical Bonner-Ebert sphere mass ($M_{BE}$) 
    for different Mach numbers (see text for definitions). 
 The dashed line shows
$M_\mathrm{core}=M_{BE}$ and the shaded area marks the region with 
$0.5<M_\mathrm{core}/M_{BE}<2$. Right: $M_\mathrm{core}$ versus ${\cal
M}$. The dashed line shows a fit of the median $M_\mathrm{core}$ 
with ${\cal M}$, giving $M_\mathrm{core}/M_0=0.097{\cal
M}^{-0.23}$.}
\label{fig:mBE_Mach_mcore}
\end{figure*}

The median $t_\mathrm{coll}$-core sizes
and masses are quite similar for different Mach numbers, as a result of
the similar post-shock density across different Mach numbers
$\rho_\mathrm{post-shock}/\rho_0 \sim 10-20$.
Using the fiducial values of $M_0$ and $L_0$ from Table \ref{table:units}, the
characteristic core masses and radii are $\sim (1-2)M_\sun$ and $\sim
(0.02-0.03)$pc.
Figure \ref{fig:mBE_Mach_mcore} plots the mass defined by gravitational
potential alone of $t_\mathrm{coll}$-cores ($M_\mathrm{core}$) 
versus the post shock critical Bonner-Ebert sphere mass ($M_{BE}$) and
Mach number (${\cal M})$.
Setting $P_\mathrm{edge}=c_s^2\rho_\mathrm{post-shock}$ in Equation
(\ref{eq:M_BE_crit}), the critical Bonner-Ebert mass becomes
\begin{equation}\label{eq:M_BE}
    \frac{M_{BE}}{M_0} =
    0.22\left( \frac{\rho_\mathrm{post-shock}}{\rho_0} \right)^{-1/2}.
\end{equation}
We fit the post-shock density field
(defined as zones with $\rho>1.5\rho_0$) with
a log-normal distribution with mean value of 
$\log_{10}(\rho/ \rho_0)$ equal to $\mu$ and dispersion $\sigma$. This
distribution remains roughly constant over the time of core formation. 
We then used $\mu$ and $\sigma$ at the time when the core formation rate is
roughly at its peak to estimate the corresponding distribution of
$M_{BE}$: from Equation (\ref{eq:M_BE}), 
$M_{BE}(\rho_\mathrm{post-shock})/M_0$ also follows
a log-normal distribution with a mean value
$\log_{10}(1.2/\pi^{3/2})-0.5\mu$ and 
dispersion $0.5\sigma$, which is plotted as dots and error bars on the
x-axis in Figure \ref{fig:mBE_Mach_mcore}.
On the y-axis of Figure \ref{fig:mBE_Mach_mcore}, 
we plot the median value of $M_\mathrm{core}$ and error bar showing the median absolute
deviation of $M_\mathrm{core}$ in logarithm space, 
as listed in Table \ref{table:core_properties}.
Typically $0.5 \lesssim M_\mathrm{core}/M_{BE} \lesssim 2$, 
consistent with the general range expected for isothermal fragmentation.
The median core mass depends on inflow Mach number as
$M_\mathrm{core} \sim {\cal M}^{-0.23}$. As noted above, the weak variation in 
$\rho_\mathrm{post-shock}$, and resulting weak dependence of the characteristic
mass on ${\cal M}$, is a result of instability of the post-shock layer. In
magnetized simulations, for which these instabilities are suppressed, 
the post-shock density is close to the expectation of magnetized isothermal shock
$\rho_\mathrm{post-shock}\propto {\cal M}$. Combining this effect with
anisotropic flows along magnetic field lines leads to a stronger dependence of
core mass on Mach number close to $M_\mathrm{core} \propto {\cal M}^{-1}$
(Chen \& Ostriker 2015, in preparation).

With increasing Mach number, we find that
more cores form per unit area in each simulation run. Our fitting gives 
$n_\mathrm{core}/(L/L_0)^2 \sim {\cal M}^{0.54}$ for $t_\mathrm{coll}$-cores,
similar to what would be expected from the decreasing separation of filaments
$\propto L_0{\cal M}^{-1/2}$ of Equation (\ref{eq:lambda_m}).

In comparison to the  $t_\mathrm{coll}$-cores, 
the $t_1$-cores show a wider range of physical properties. This
can be explained by the fact that they include cores at very different 
evolutionary stages, as
discussed in detail in the following sections. The median masses of cores are
lower, and median radii are higher, for $t_1$-cores compared to
$t_\mathrm{coll}$ cores. The properties of $t_1$-cores
also show a stronger dependence on Mach number, because they represent 
the non-linear size and mass in the post-shock sheet that scales as ${\cal
M}^{-1/2}$ (see Equation (\ref{eq:lambda_m}) and (\ref{eq:M_m})), as also shown
in the earlier studies by \citetalias{GO2011}.

\begin{table*}[htbp]
    \centering
    \caption{Summary of Physical Properties\label{table:core_properties} for
    $t_\mathrm{coll}$-cores. }
    \begin{tabular}{l  l l l  l l }
        \tableline
        \tableline
        Model &
        \multicolumn{1}{l}{\tablenotemark{a}$\log_{10}(M_\mathrm{core}/M_0)$}
        &\multicolumn{1}{l}{\tablenotemark{a}$\log_{10}(r_\mathrm{core}/L_0)$}
        &\multicolumn{1}{l}{\tablenotemark{a,b}$\log_{10}(\rho_\mathrm{core}/\rho_0)$}
        & $n_\mathrm{core,total}$
        &\multicolumn{1}{l}{\tablenotemark{c}$n_\mathrm{core}/(L/L_0)^2$} \\
        \tableline
        \textbf{M02L6N256  }&$\mathbf{-1.07\pm0.29}$&$\mathbf{-1.47\pm0.33}$
        &$\mathbf{2.70\pm0.66}$&\textbf{267   }&\textbf{2.47 }\\
        M02L4N256  &$-1.08\pm0.27$&$-1.45\pm0.28$
        &$2.63\pm0.55$&239   &2.49 \\
        M02L4N128  &$-0.99\pm0.17$&$-1.41\pm0.21$
        &$2.64\pm0.44$&169   &2.64 \\
        M02L4N64   &$-0.82\pm0.14$&$-1.23\pm0.14$
        &$2.27\pm0.29$&220   &1.72 \\
        \tableline
        \textbf{M04L6N256  }&$\mathbf{-1.09\pm0.22}$&$\mathbf{-1.42\pm0.21}$
        &$\mathbf{2.56\pm0.42}$&\textbf{346   }&\textbf{4.81 }\\
        M04L4N256  &$-1.12\pm0.21$&$-1.46\pm0.21$
        &$2.67\pm0.45$&266   &5.54 \\
        M04L4N128  &$-1.08\pm0.15$&$-1.41\pm0.17$
        &$2.58\pm0.36$&264   &5.50 \\
        \tableline
        \textbf{M08L2N512  }&$\mathbf{-1.18\pm0.28}$&$\mathbf{-1.55\pm0.29}$
        &$\mathbf{2.84\pm0.60}$&\textbf{232   }&\textbf{7.25 }\\
        M08L2N256  &$-1.17\pm0.18$&$-1.51\pm0.19$
        &$2.70\pm0.42$&197   &8.21 \\
        M08L2N128  &$-1.15\pm0.14$&$-1.49\pm0.14$
        &$2.70\pm0.32$&182   &9.10 \\
        M08L4N256  &$-1.17\pm0.19$&$-1.50\pm0.20$
        &$2.70\pm0.44$&121   &7.56 \\
        \tableline
        \textbf{M16L2N512  }&$\mathbf{-1.26\pm0.20}$&$\mathbf{-1.57\pm0.21}$
        &$\mathbf{2.87\pm0.48}$&\textbf{122   }&\textbf{7.62 }\\
        M16L1.5N512&$-1.24\pm0.23$&$-1.53\pm0.25$
        &$2.71\pm0.51$&107   &7.93 \\
        M16L2N256  &$-1.33\pm0.17$&$-1.62\pm0.19$
        &$2.91\pm0.42$&158   &13.17\\
        \tableline
    \end{tabular}
    \tablenotetext{1}{Median value $\pm$ median absolute deviation (MAD)}
    \tablenotetext{2}{
    $\rho_\mathrm{core} \equiv M_\mathrm{core}$/$V_\mathrm{core}$ for each core.}
    \tablenotetext{3}{Average number of cores per simulation run per $L_0^2$
    area in the $x-y$ plane.}
\end{table*}

\begin{table*}[htbp]
    \centering
    \caption{Summary of Physical Properties\label{table:core_properties_t0} for
    $t_1$-cores. }
    \begin{tabular}{l  l l l  l l }
        \tableline
        \tableline
        Model &
        $\log_{10}(M_\mathrm{core}/M_0)$ &$\log_{10}(r_\mathrm{core}/r_0)$
        &$\log_{10}(\rho_\mathrm{core}/\rho_0)$ & $n_\mathrm{core,total}$
        &$n_\mathrm{core}/(L/L_0)^2$ \\
        \tableline
        \textbf{M02L6N256}&$\mathbf{-1.33\pm0.38}$&$\mathbf{-1.16\pm0.15}$
        &$\mathbf{1.53\pm0.09}$&\textbf{94    }&\textbf{0.87 }\\
        M02L4N256  &$-1.44\pm0.35$&$-1.24\pm0.12$
        &$1.59\pm0.14$&86    &0.90 \\
        \tableline
        \textbf{M04L6N256}&$\mathbf{-1.51\pm0.24}$&$\mathbf{-1.31\pm0.09}$
        &$\mathbf{1.83\pm0.12}$&\textbf{151   }&\textbf{2.10 }\\
        M04L4N256  &$-1.43\pm0.27$&$-1.32\pm0.10$
        &$1.82\pm0.16$&103   &2.15 \\
        \tableline
        \textbf{M08L2N512}&$\mathbf{-1.57\pm0.32}$&$\mathbf{-1.44\pm0.13}$
        &$\mathbf{2.00\pm0.17}$&\textbf{139   }&\textbf{4.34 }\\
        \tableline
        \textbf{M16L2N512}&$\mathbf{-1.60\pm0.29}$&$\mathbf{-1.51\pm0.18}$
        &$\mathbf{2.36\pm0.29}$&\textbf{103   }&\textbf{6.44 }\\
        M16L1.5N512&$-1.46\pm0.28$&$-1.48\pm0.18$
        &$2.30\pm0.26$&83    &6.15 \\
        \tableline
    \end{tabular}
    \tablecomments{Notations same as Table \ref{table:core_properties}.}
\end{table*}

\begin{figure*}[htbp]
\centering
\includegraphics[width=0.8\linewidth]{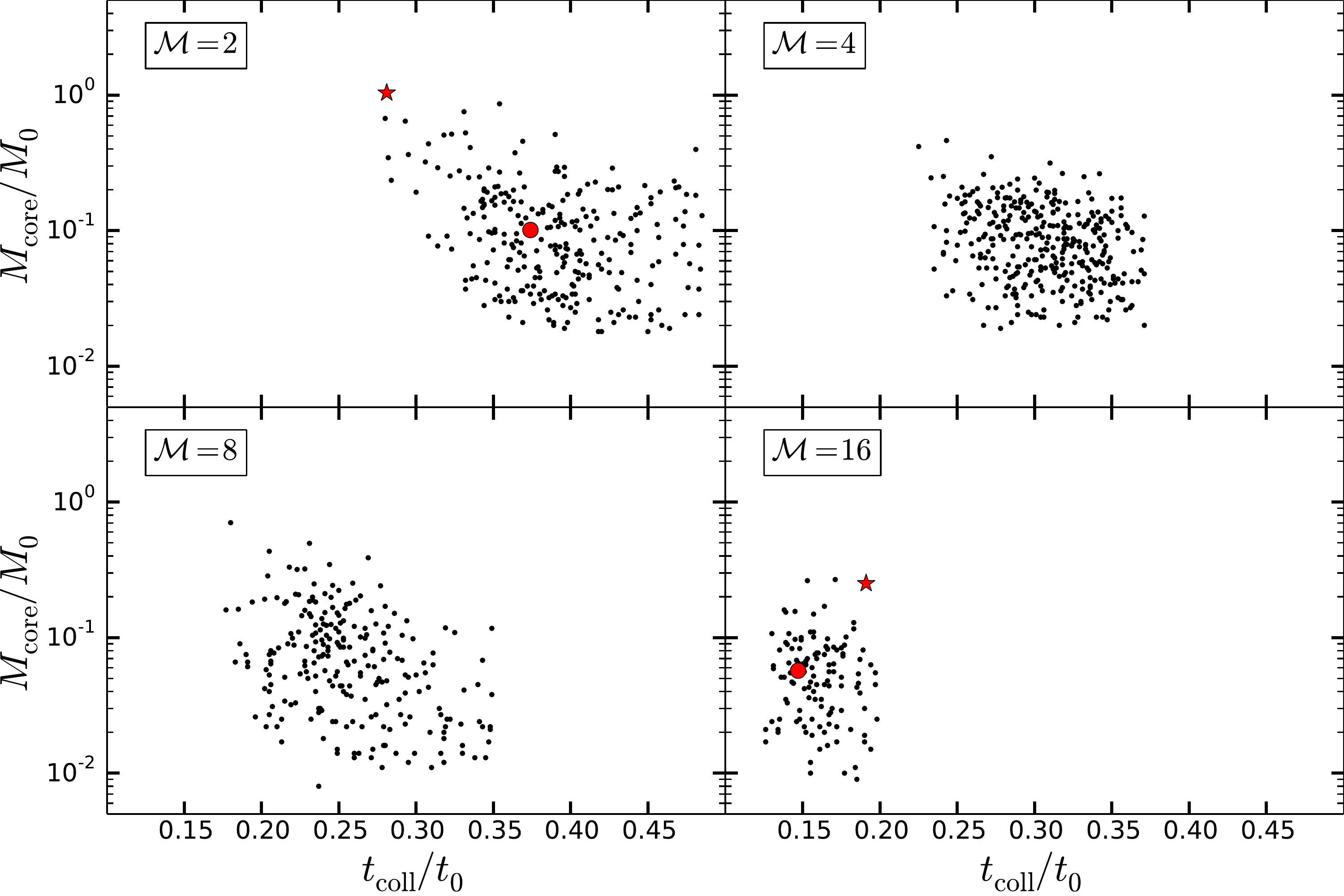}
\caption{Mass and time of collapse of individual $t_\mathrm{coll}$-cores 
for different ${\cal M}$. The red dots and
stars mark a typical mass core and a high mass core in ${\cal M}=2,16$
simulations as illustrated in Figure \ref{fig:corefind_tcoll}.}
\label{fig:mcore_ti}
\end{figure*}

Figure \ref{fig:mcore_ti} plots the mass and time of collapse for
every $t_\mathrm{coll}$-core. For higher Mach numbers, the cores form sooner
and in a less extended period of time, as explained in \S \ref{section:non-linear}. 
This also leads to more similarity between $t_1$ and
$t_\mathrm{coll}$ cores (see Table \ref{table:core_properties} and
\ref{table:core_properties_t0}) for higher Mach numbers.
There is a similar pattern in models with different 
${\cal M}$. The average core mass decreases slightly with time, as smaller
cores form in dense filaments which become more developed at later times.
At the end of each simulation, the core formation rate drops, when most
material in the post-shock region has been accreted by the sink particles. This is
further discussed in \S \ref{section:SFR}.

The median value and the median absolute deviation at a logarithm scale of
core mass and radius in high-amplitude velocity perturbation simulations are
$\log_{10}(M_\mathrm{core}/M_0) = -1.29\pm0.22$ and 
$\log_{10}(r_\mathrm{core}/L_0)=-1.86\pm0.18$.
This is about $\sim 60$\% of the core mass and radius in low-amplitude
perturbation cases (see the M08L2N256 model in Table
\ref{table:core_properties}), but subject to the limited statistics of 21 cores.
There are also a few massive turbulence supported
cores formed in high-amplitude simulations, which are not seen in low-amplitude cases;
this is further discussed in the last paragraph of \S \ref{section:CMF}


\subsubsection{Core Binding}\label{section:binding}

\begin{figure}[htbp]
\centering
\includegraphics[width=\linewidth]{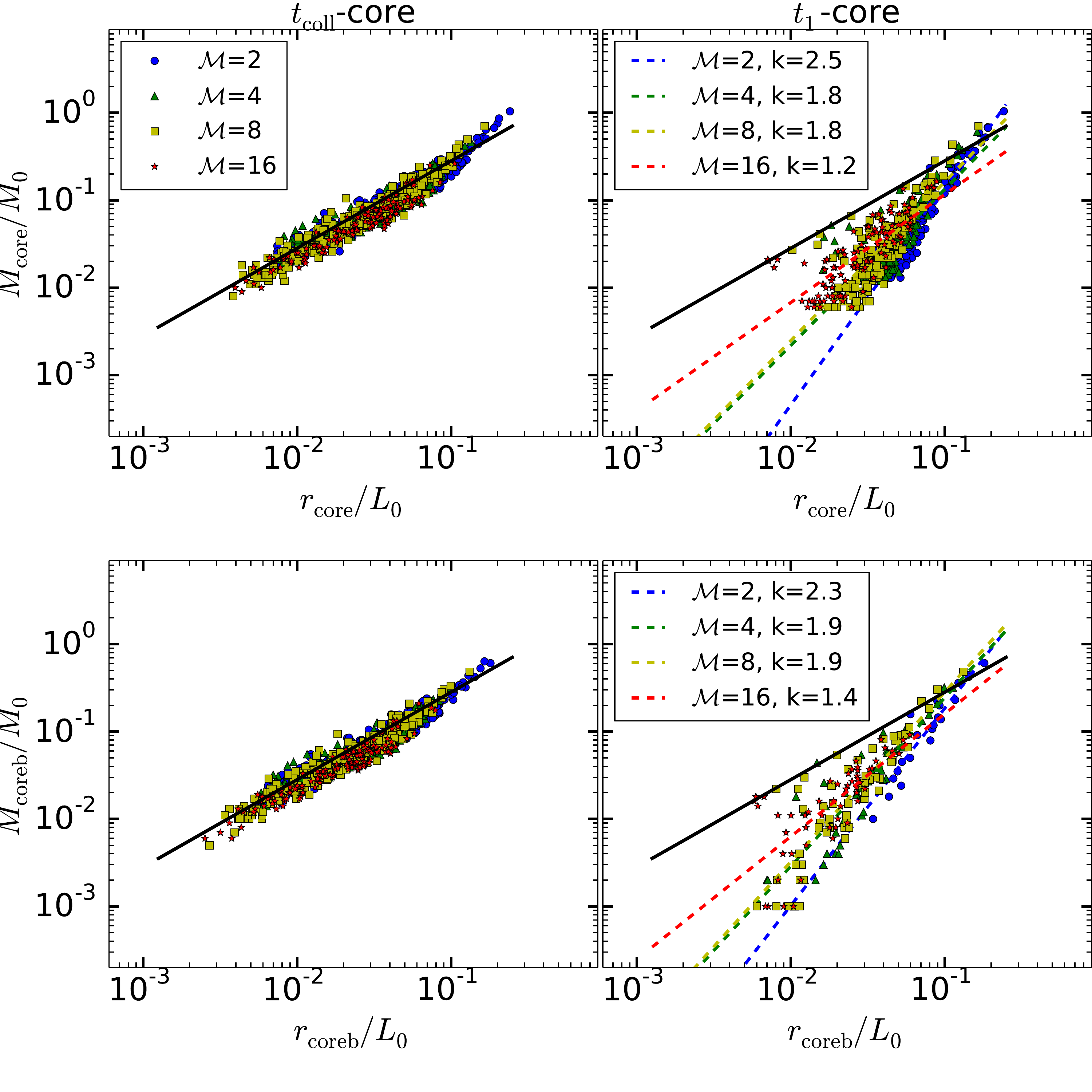}
\caption{
Mass-radius relation of cores.
Left panels: $t_\mathrm{coll}$-cores defined by gravitational potential
alone (upper) and bound $t_\mathrm{coll}$-cores with 
$E_G+E_\mathrm{th}<0$ (lower). The solid line plots the mass-radius relation
for the LP density profile in Equation (\ref{eq:M_LP}). 
Right panels: same as for left, but for $t_1$-cores. 
Dotted lines plot fits of mass-radius relation $M\propto r^k$ 
for different ${\cal M}$, and the values of fitted $k$ are listed
on the plot legends. The $t_1$-cores are less internally stratified than
$t_\mathrm{coll}$-cores.
}
\label{fig:mcore_rcore}
\end{figure}

\begin{figure}[htbp]
\centering
\includegraphics[width=\linewidth]{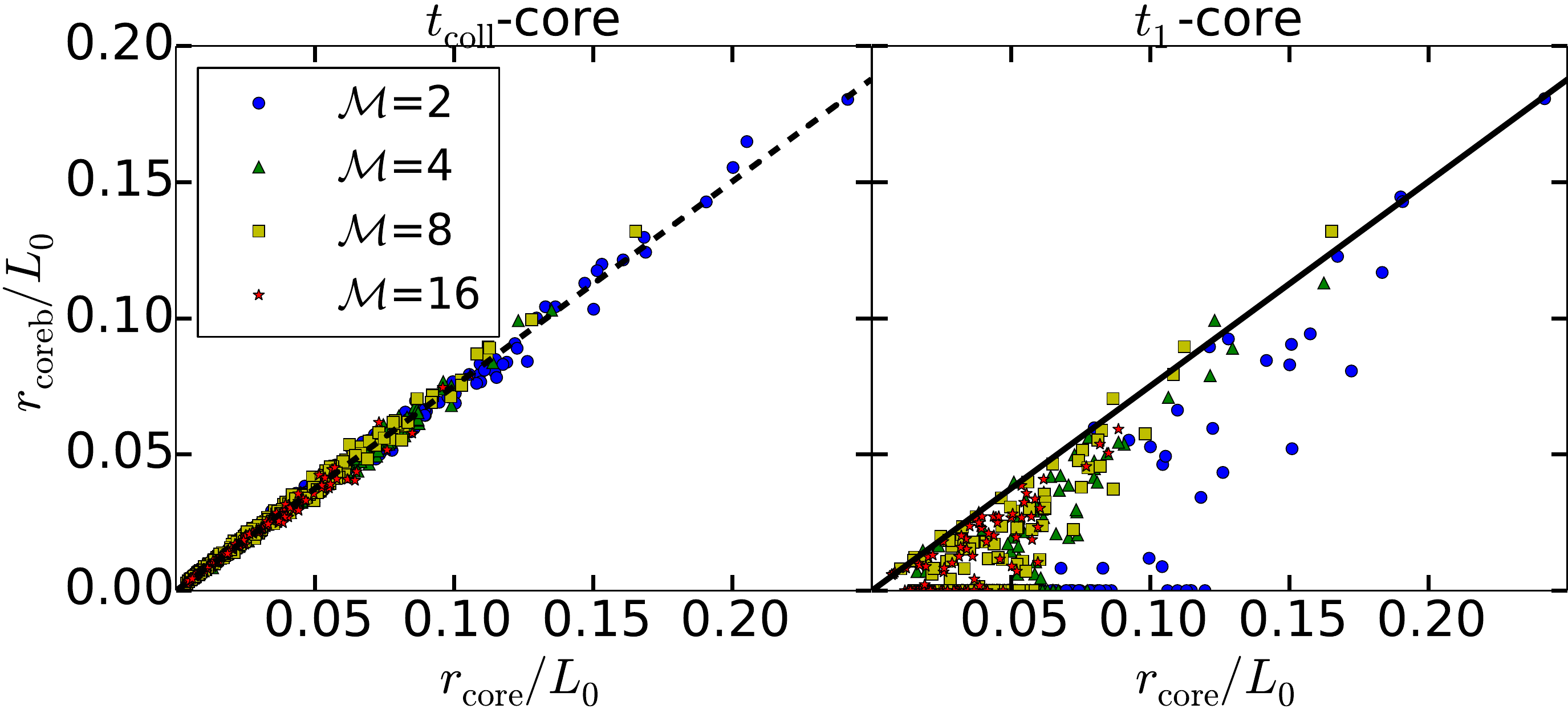}
\caption{
    Radius of cores defined by gravitational potential alone ($r_\mathrm{core}$) 
    compared to radius of the portion of the bound
    cores with $E_G+E_\mathrm{th}<0$ ($r_\mathrm{coreb}$). 
    Left: $t_\mathrm{coll}$-cores. The dashed
line gives a linear fit of $r_\mathrm{coreb}/r_\mathrm{core} = 0.75$. 
Right: $t_1$-cores.
The solid line indicates $r_\mathrm{coreb}/r_\mathrm{core} = 0.75$. Most
cores at $t_1$ are not strongly bound.}
\label{fig:rcore_rcoreb}
\end{figure}

Since the $t_\mathrm{coll}$-cores are identified at the time of collapse by
definition, they are expected to have density profiles close to the LP profile.
This is indeed the case, as shown below  (see also Figure
\ref{fig:core_propertiesM02} and \ref{fig:core_propertiesM16}).
Integrating Equation (\ref{eq:rho_LP}), the
mass of the core will be proportional to its radius 
\begin{equation}\label{eq:M_LP}
    \frac{M_\mathrm{LP}(r)}{M_0} = \frac{8.86}{\pi} \frac{r}{L_0}.
\end{equation}
Solving the Poisson equation
\begin{equation}
    \frac{1}{r^2} \frac{\di}{\di r} \left( r^2 \frac{\di \Phi}{\di r} \right)
    = 4\pi G \rho_\mathrm{LP}(r),
\end{equation}
the gravitational potential of an isolated core with LP profile can be written
as
\begin{equation}
    -\Delta \Phi(r) \equiv \Phi_\mathrm{edge} - \Phi(r) 
      = -8.86c_s^2 \ln \frac{r}{r_\mathrm{core}}
\end{equation}
where $\Phi_\mathrm{edge}$ is the gravitational potential
at the edge of the core, $r_\mathrm{core}$.
Using $c_s^2= -\Delta \Phi(r_\mathrm{coreb})$ for the definition of bound cores, 
we can obtain the relationship of $r_\mathrm{coreb}$ and $r_\mathrm{core}$:
\begin{equation}\label{eq:r_coreb}
    r_\mathrm{coreb} = r_\mathrm{core}e^{-1/8.86} \approx 0.9 r_\mathrm{core}.
\end{equation}
The left panels of Figure \ref{fig:mcore_rcore} shows the mass-radius 
relation of the $t_\mathrm{coll}$-cores defined by gravitational potential alone and the
bound $t_\mathrm{coll}$-cores with $E_G+E_\mathrm{th}<0$. 
The results are consistent with equation (\ref{eq:M_LP}), i.e. 
$M_\mathrm{core}\propto r_\mathrm{core}$ 
and $M_\mathrm{coreb}\propto r_\mathrm{coreb}$, showing that essentially all
cores approach the LP density profile at the time of collapse. The left panel of 
Figure \ref{fig:rcore_rcoreb} plots the size of every 
$t_\mathrm{coll}$-core defined by gravitational
potential alone compared to the corresponding bound core with
$E_G+E_\mathrm{th}<0$, showing that $r_\mathrm{coreb} \propto r_\mathrm{core}$,
as expected from Equation (\ref{eq:r_coreb}).
Due to the potential of surrounding cores and filaments,
gravitational potential profiles at the edges of the $t_\mathrm{coll}$-cores in
the simulations are slightly flatter than that they would be for isolated cores.
Therefore, we find $r_\mathrm{coreb}/r_\mathrm{core} \approx 0.75$, 
slightly smaller than the coefficient in Equation (\ref{eq:r_coreb}).

In comparison, many $t_1$-cores are not strongly centrally concentrated
or gravitationally bound, as shown in the right panels of Figure \ref{fig:mcore_rcore} 
and Figure \ref{fig:rcore_rcoreb}. The mass radius relation of $t_1$-cores is
found to be $M \propto r^k$ with $k=1.2-2.5$, implying that they have flatter
density profiles than the LP profile or critical Bonner-Ebert sphere. 
This is consistent with results for $t_1$-cores in previous
simulations with or without magnetic fields \citep{GO2009, CO2014}, and similar
to many core-property surveys in different molecular clouds \citep[e.g.,][]{
CR2010, Kirk2013}, in which $k=1.4-2.4$ with various
tracers (see Figure 7 and
corresponding discussions in \citet{Kirk2013}). This suggests that cores
identified at any given time in simulations or observations ($t_1$-cores in our
case) can be in very different stages of evolution. Because any individual core 
spends most of its lifetime in the core building stage, many of the cores in a
given snapshot of time will resemble sub-critical Bonner-Ebert
spheres\footnote{Cores at very early building stages would, however, be
difficult to pick out in a noisy background, as the density contrast would be
low.}, 
with only a fraction of them more evolved and in the core collapse stage.
Some of the cores we identify, especially low mass cores, are 
structures with weak self-gravity and may not finally collapse. The spread
of mass-radius relation is larger for lower mass cores in the right panels of
Figure \ref{fig:mcore_rcore}, implying a wider range of evolutionary stages.
Interestingly, we find a shallower mass-radius relation at $t_1$ 
for higher Mach number simulations, 
indicating those cores are more evolved and therefore more centrally concentrated.
With complete population studies from {\sl Herschel} and {\sl ALMA} that are able to quantify the
relative populations of strongly concentrated cores versus those with flatter
density profiles, the timescales for core collapse versus core building stages
can be measured (see \citet{Marsh2014} for initial Herschel results).

\begin{figure*}[htbp]
\centering
\includegraphics[width=0.8\linewidth]{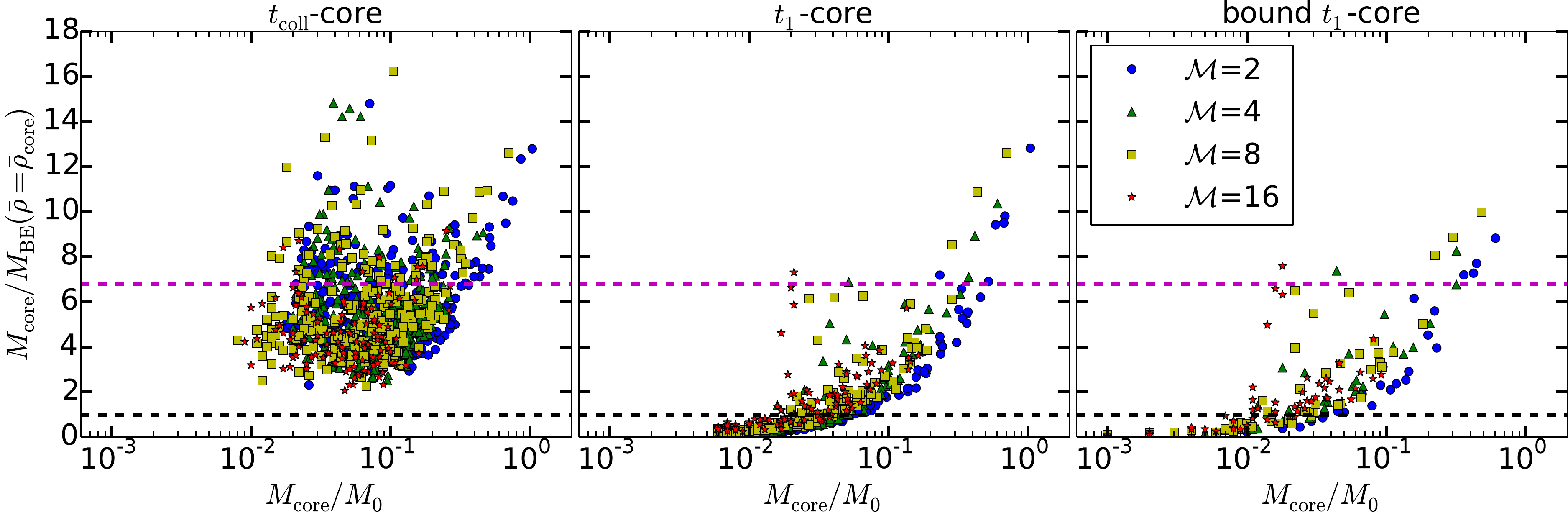}
\caption{The ratio of core mass to the mass of a critical Bonner-Ebert sphere with the
    same internal average density, versus core mass. 
    Left:$t_\mathrm{coll}$-cores defined by the gravitational potential
alone. Middle and right: $t_1$-cores, defined by gravitational potential
alone (middle), and bound cores with $E_G+E_\mathrm{th}<0$ (right).
The magenta and black dashed lines are
$M_\mathrm{core}/M_{BE}(\bar{\rho}=\bar{\rho}_\mathrm{core}) = 6.78$ 
(for LP profile), and $1$, respectively.}
\label{fig:Mcore_MBE}
\end{figure*}

Another way to quantify whether a core is gravitationally unstable or not
is to look at the ratio of core mass to the critical Bonner-Ebert sphere
with the same average density of the core.
For the LP density profile, the ratio
$M_\mathrm{core}/M_{BE}(\bar{\rho}=\bar{\rho}_\mathrm{core})=6.8$ (see Equation
(\ref{eq:M_BE_crit})). 
In Figure
\ref{fig:Mcore_MBE}, all $t_\mathrm{coll}$-cores are gravitationally unstable,
with the ratio
$M_\mathrm{core}/M_{BE}(\bar{\rho}=\bar{\rho}_\mathrm{core})>1$. 
In comparison, among all the $t_1$-cores
, only $\sim45\%$ of cores defined by the gravitational potential
alone, and $\sim60\%$ of bound cores with $E_G+E_\mathrm{th}<0$, are
gravitationally unstable with 
$M_\mathrm{core}/M_{BE}(\bar{\rho}=\bar{\rho}_\mathrm{core})>1$. 
There is also a clear trend
that this ratio increases with mass, which is not present for the
$t_\mathrm{coll}$-cores. This again suggests that many
$t_1$-cores, especially the low mass ones  ($M_\mathrm{core}/M_0 \lesssim 10^{-2}$
in the middle and right panels of Figure \ref{fig:Mcore_MBE}), 
 are transient structures from 
turbulence perturbations, and may not finally give rise to star formation. 
This is also evident in the CMF described in \S \ref{section:CMF}.

The core binding properties in high-amplitude perturbation simulations are
similar to the low-amplitude cases as discussed above. All of the
$t_\mathrm{coll}$-cores are strongly gravitationally bound, whereas most of the
$t_1$-cores, especially the lower mass ones, are not strongly concentrated or
highly stratified.

\subsubsection{Core Shape}

\begin{figure}[htbp]
\centering
\includegraphics[width=\linewidth]{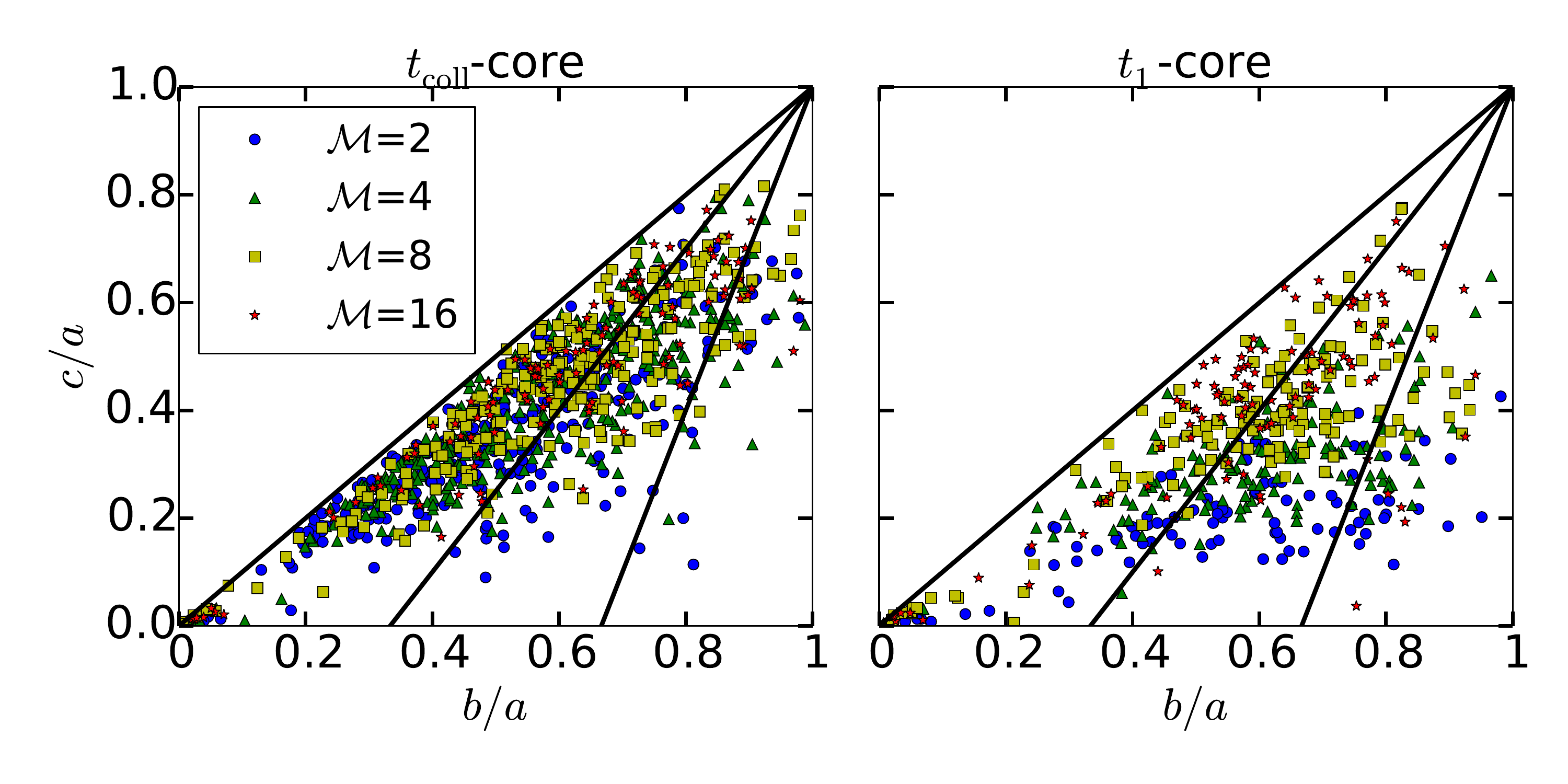}
\caption{Distributions of three-dimensional core aspect ratios of cores defined
    by gravitational potential alone for each Mach number. See also Figure 19
    in \citetalias{GO2011}. Cores lying on $c/a=b/a$ are formally prolate and along
    $b/a = 1$ are formally oblate. We subdivide (see diagonal lines) and classify as
    follows: approximately prolate (between $c/a=1$ and $c/a=1.5b/a-0.5$),
    triaxial(between $c/a=1.5b/a-0.5$ and $c/a=3b/a-2$), and approximately
    oblate(between $c/a=3b/a-2$ and $b/a=1$). Left:
    $t_\mathrm{coll}$-cores. Approximately 66\% of $t_\mathrm{coll}$-cores are
    prolate, 29\% triaxial and 5\% oblate, independent of
    Mach number. Right: $t_1$-cores. Approximately $30-50\%$ of
$t_1$-cores are prolate, $30-40\%$ triaxial, and $10-20\%$ 
oblate. Slightly more cores are prolate for higher Mach numbers.}
\label{fig:vratio}
\end{figure} 

The shape of a core can be described by the eigenvalues of the moment of
inertia tensor $I_{ij}\equiv \int \rho x_i x_j \di^3 \mathbf{x}$
\citep[e.g.,][]{Gammie2003, NL2008}), or the covariance matrix of offsets with respect
to the center of mass $M_{ij} \equiv \int\rho x_i \rho x_j \di^3 \mathbf{x}$
\citepalias{GO2011}.
Let a,b and c be the lengths of the principal axes
and $a \geq b \geq c$. A prolate core has $b/a=c/a$, and an oblate core 
$b/a=1$. The core aspect ratio distribution in Figure \ref{fig:vratio}
shows several interesting features. First, most cores are prolate
and triaxial, and only a small fraction are oblate. If cores form via
fragmentation in isothermal filaments, they are expected to be initially prolate, since
the fragmentation scale along filaments is expected to 
have separations larger than the filament diameter \citep{Nagasawa1987, IM1992}.
In fact, the typical $t_1$-core aspect ratios are smaller than the ratio of spacing to
diameter for the fastest growing mode of filament fragmentation, 
again suggesting that cores are not
exclusively formed as instabilities in filaments.
Second, the fraction of cores that are prolate is higher
in $t_\mathrm{coll}$-cores than $t_1$-cores, implying more
evolved cores tends to be more prolate, whereas cores formed via gravitational
instability of filaments become more spherical in time. Last, among all $t_1$-cores, 
more are prolate for higher Mach numbers, consistent with
those cores in high Mach number simulations being slightly more evolved,
as also indicated in the left panels of Figure \ref{fig:mcore_rcore}.

\subsubsection{Core Mass Function}\label{section:CMF}
\begin{figure*}[htbp]
\centering
\includegraphics[width=0.9\linewidth]{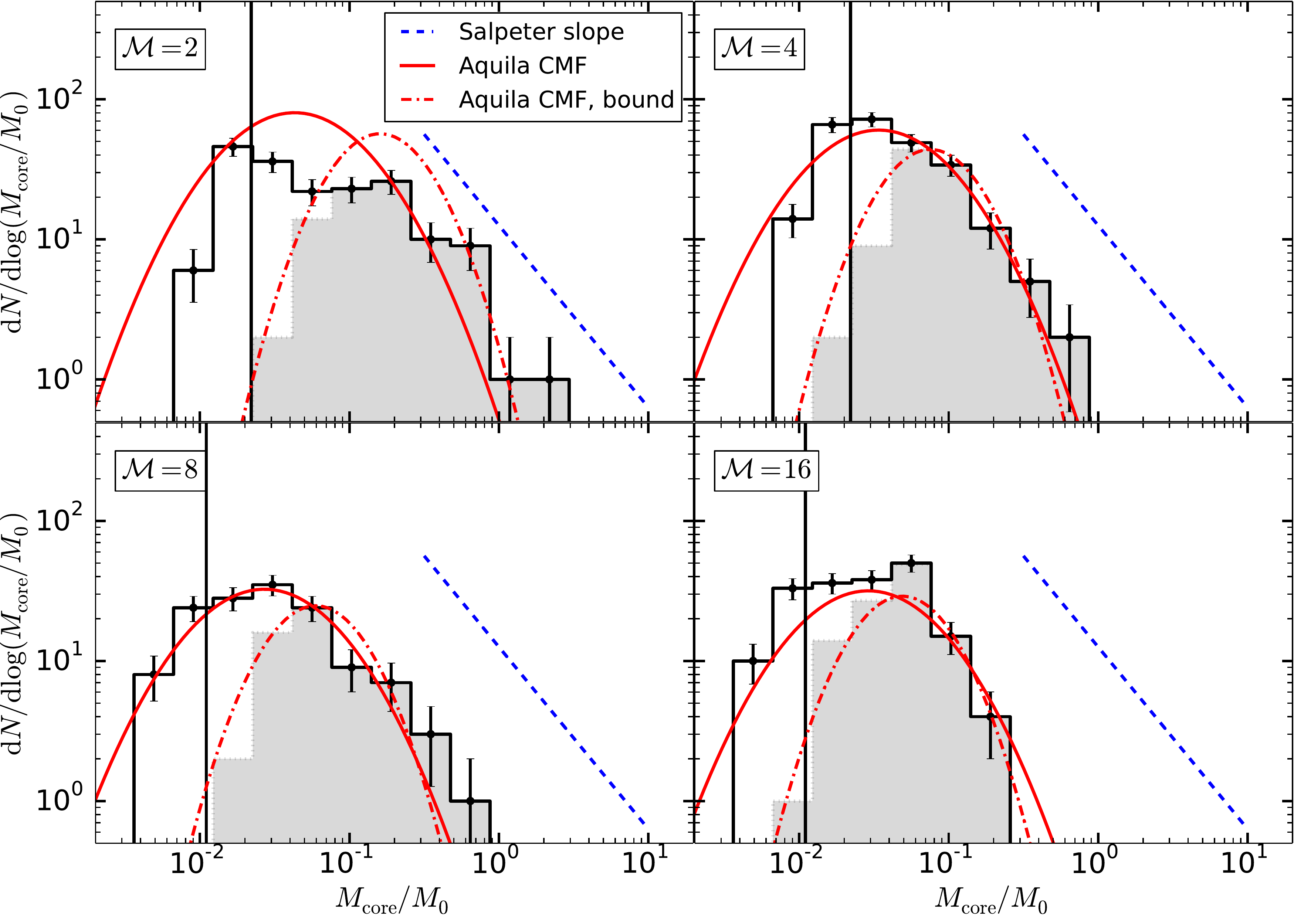}
\caption{{\footnotesize Core mass function (CMF) of $t_1$-cores 
from different Mach number simulations. 
To improve statistics, models of M02L4N256, M04L4N256,
M16L1.5N512 are included in addition to the bold-face models in Table
\ref{table:model_parameters}. In each panel, 
the histogram shows the CMF of all $t_1$-cores, 
with the shaded area denoting the gravitationally bound portion
(see Figure \ref{fig:Mcore_MBE}).  The black vertical line
marks the resolution limit similar to Figure \ref{fig:convergence_Mcore}. 
The red solid and dash-dotted curves plot the fit to the observed CMF 
(log-normal distribution with standard deviation of 0.43 and 0.30)
of the Aquila region of the full sample and the gravitationally bound
portion of cores \citep{Andre2010, Konyves2010}, 
shifted horizontally to match the position of median core mass in the
corresponding histogram, and normalized so that the total mass $>0.01M_0$ 
(similar to their observational resolution limit) is the same as in the
histogram.
The median core mass for the full sample and the gravitationally
bound portion of $t_1$-cores are $M_\mathrm{core}/M_\sun =$1.0, 0.8, 0.6, and 0.7,
and  $M_\mathrm{core}/M_\sun =$3.7, 1.7, 1.4 and 1.1 for ${\cal M}=$2, 4, 8,
and 16, using $M_0=23M_\sun$ in Table \ref{table:units}. The blue dashed line
segment shows the Salpeter slope $\di N/\di M \propto M^{-2.3}$, for reference.
The error bars show $\sqrt{N}$, where $N$ is the number
of cores in each bin.}}
\label{fig:CMF_t1}
\end{figure*}

\begin{figure*}[tbp]
\centering
\includegraphics[width=0.9\linewidth]{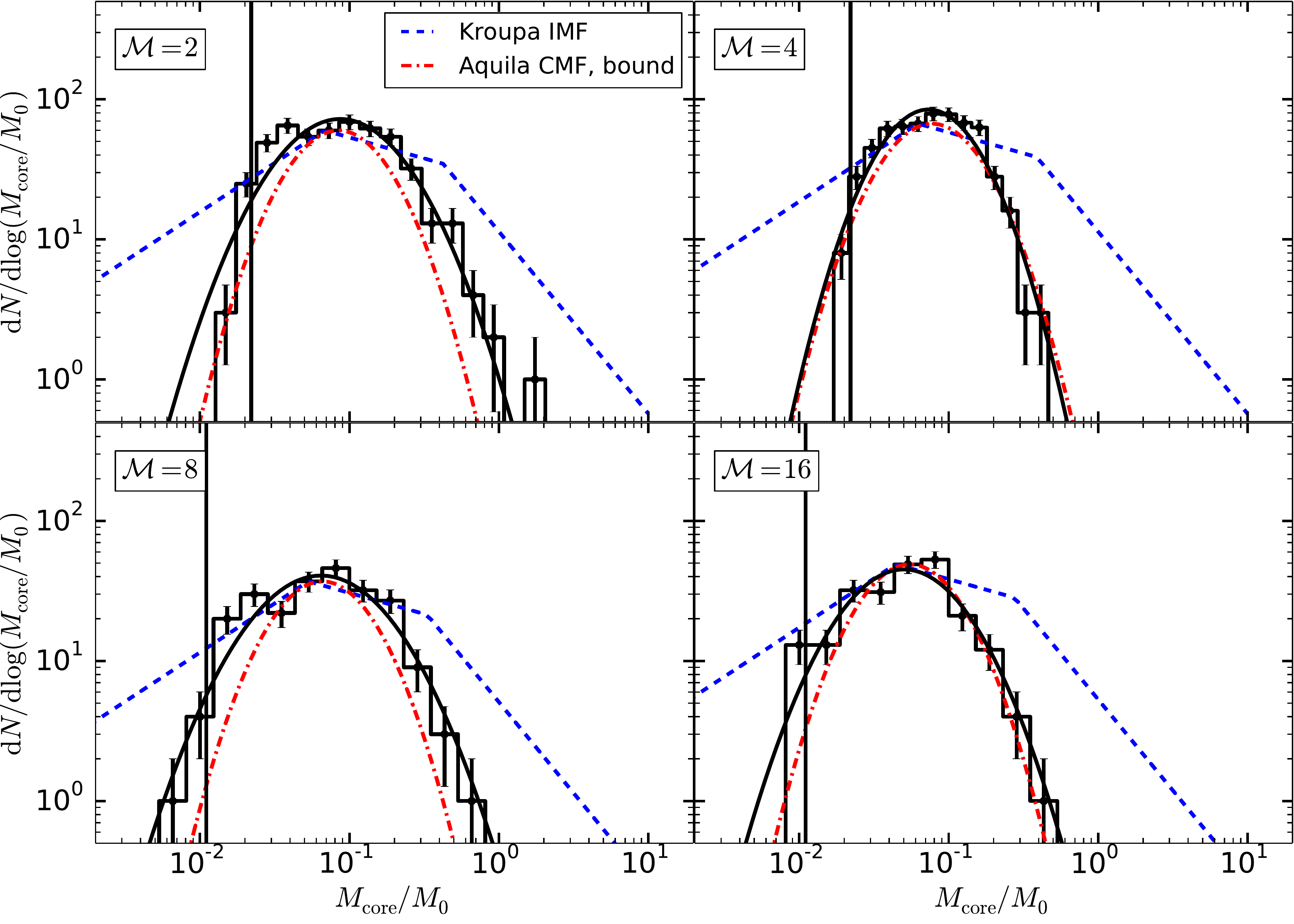}
\caption{Histogram shows core mass function (CMF) of $t_\mathrm{coll}$-cores, similar to Figure
    \ref{fig:CMF_t1}. In each panel, the black solid curve shows the log-normal
    fit to the CMF, giving standard deviations of 0.29, 0.36, 0.39 and 0.35 for 
    ${\cal M}=$2, 4, 8, and 16. The median $t_\mathrm{coll}$-core masses 
are $M_\mathrm{core}/M_\sun =$1.9, 1.8, 1.5 and 1.3,
for ${\cal M}=$2, 4, 8, and 16, using $M_0=23M_\sun$ in Table
\ref{table:units}. Also shown in blue dashed curve and red dash-dotted curve, 
for reference, is the Kroupa IMF \citep{Kroupa2001} 
and the bound core CMF fit from Aquila \citep{Konyves2010}, both
shifted horizontally to match median masses.
}
\label{fig:CMF_tcoll}
\end{figure*}

\begin{figure*}[htbp]
\centering
\includegraphics[width=0.9\linewidth]{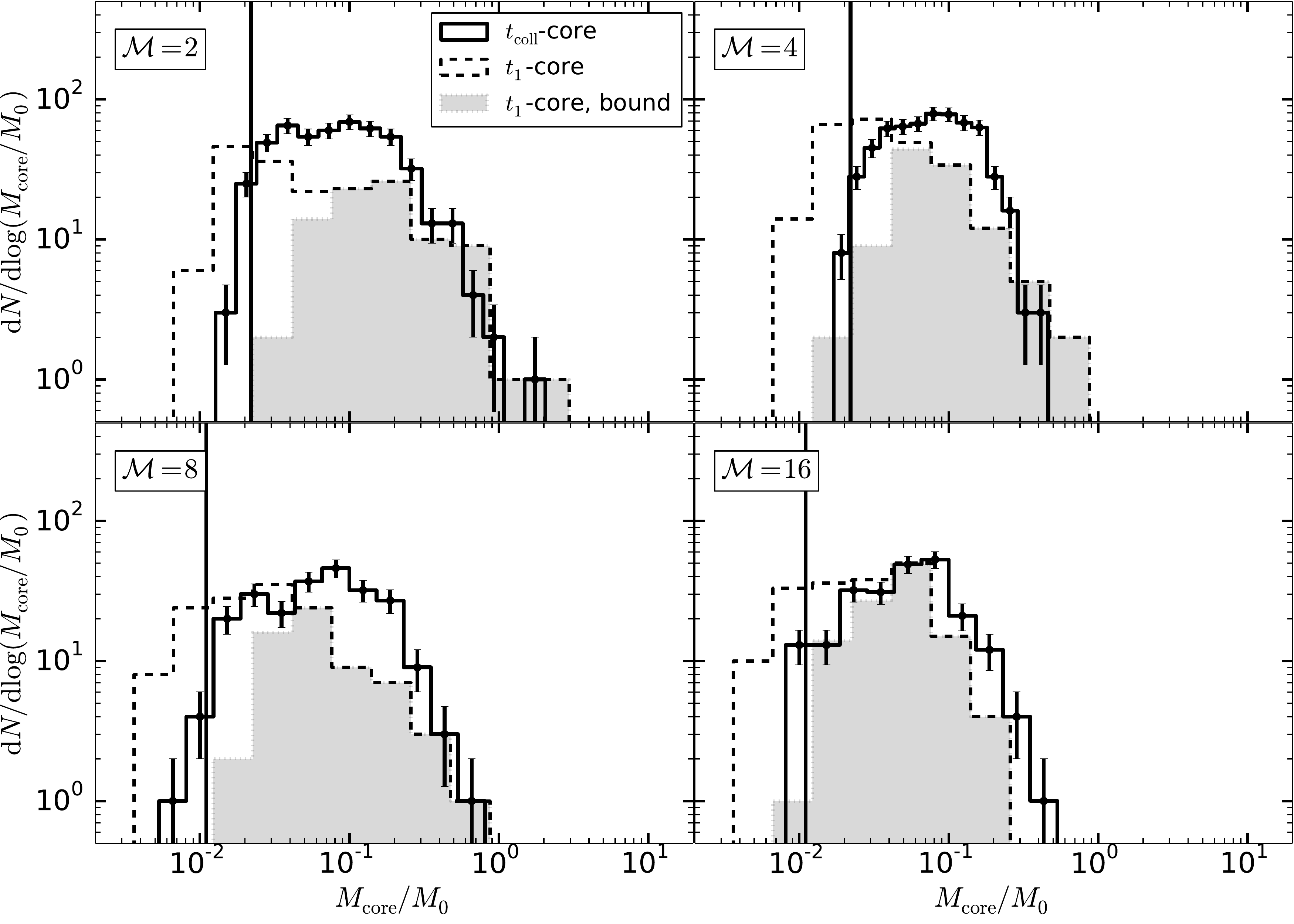}
\caption{Comparison of CMFs of $t_1$-cores and $t_\mathrm{coll}$-cores (see
    also Figure \ref{fig:CMF_t1} and \ref{fig:CMF_tcoll}). In each panel, the
    solid histogram plots the CMF of $t_\mathrm{coll}$-cores, with error bars
    showing $\sqrt{N}$, where $N$ is the number of cores in each bin. The
    dashed histogram shows the CMF of $t_1$-cores, and the shaded part the
    gravitationally bound portion.
}
\label{fig:CMF_simulation}
\end{figure*}

The core mass functions (CMFs) of $t_1$-cores 
are shown in Figure \ref{fig:CMF_t1}. To compare the
CMFs from our simulations to the observed CMF,we match the median mass, 
and only compare the shape of different distributions with a logarithmic horizontal
scale. This is because the mass scale $M_0$ in our simulations can be adjusted
by changing $\rho_0$ (see Equation (\ref{eq:M_J})). Figure \ref{fig:CMF_t1}
includes the (shifted) fit to full Aquila core sample of \citet{Andre2010},
and the (shifted) fit to the gravitationally bound portion \citep{Konyves2010}.
Although our statistics are limited for high-mass cores, for reference in
Figure \ref{fig:CMF_t1} we also show the slope of the Salpeter mass function
$\di N/\di M \propto M^{-2.3}$.

The CMF of the full sample of $t_1$-cores in Figure \ref{fig:CMF_t1} is
similar to the shape of the observed CMF near the peak. 
Among the $t_1$-cores, the majority of the small
cores are weakly self-gravitating transient structures. 
If we only include the gravitationally bound $t_1$-cores
(shaded histogram in Figure \ref{fig:CMF_t1}), the distribution is
similar to the gravitationally bound cores observed
in Aquila \citep{Konyves2010}. Although statistics for our simulations (and
observed) cores are too poor to fit a high-end mass slope with confidence, the
distribution of high-mass $t_1$-cores appear to be consistent with the
log-normal fit of observed cores, and shows no evidence of an extended power-law tail.

Figure \ref{fig:CMF_tcoll} shows the mass distribution of
$t_\mathrm{coll}$-cores. We also show, for each model, a log-normal fit to the
distribution. In addition, Figure \ref{fig:CMF_tcoll} includes for comparison
the bound core Aquila fit and the Kroupa IMF (shifted so that median masses
match). 
As can be seen in Figure \ref{fig:CMF_tcoll}, the CMFs of
$t_\mathrm{coll}$-cores 
show a statistically significant
($\gtrsim 3\sigma$ in the high mass bins) deficit of  high mass cores
compared to the shape of the observed IMF, and  
can be well-fitted by relatively narrow log-normal
distributions with standard deviations of $\sim 0.3-0.4$. This is simliar to
the bound core CMF log-normal fit in the Aquila cloud, which has a standard deviation
$0.3$.

Figure \ref{fig:CMF_simulation} shows a comparison between the $t_1$-cores and the
$t_\mathrm{coll}$-cores, for each value of ${\cal M}$.
It is interesting that 
the CMFs of $t_1$-cores and $t_\mathrm{coll}$-cores do not match in our
simulations. The $t_1$-cores show a wider distribution with more low-mass cores,
and the characteristic core  mass is a factor of $\sim 2$
smaller than that of $t_\mathrm{coll}$-cores. However, the bound portion of the
$t_1$-core distribution is more similar to the $t_\mathrm{coll}$-cores.

The differences between $t_1$-cores and $t_\mathrm{coll}$-cores has interesting
implications for interpreting observations.
Observations
of CMFs represent a snapshot in time of GMCs, and often do not distinguish between
gravitationally stable (weakly bound) or unstable (strongly bound) cores 
\citep[e.g.,][]{NW2007, Andre2010}. 
Our results suggest that if the cores
defined in observations are comparable to our $t_1$-cores, then 
a large fraction of the low mass cores in observed CMFs may not
end up collapsing, or gain a significant fraction of mass before they collapse.
This also implies that the observed similarities between the
CMF and IMF cannot be simply interpreted as a fundamental
correlation between stellar mass and core mass, with each core contributing the
same proportion of its mass to the final star, because we know that there is
not a one-to-one mapping between cores observed at a given instant and cores at
the time of collapse of star formation. However, the similarity between our
bound $t_1$-cores and our $t_\mathrm{coll}$-cores suggests that an analogous
selection of just gravitationally bound cores can be obtained in observations, 
to probe the CMFs that better represent the initial mass reservoir for star formation.

The difference between the CMFs of $t_\mathrm{coll}$-cores in our models 
and the observed IMF
could have several possible reasons.
First, the idealizations or physics in our simulations
may be unfavorable for high mass core formation. For example, the presence of
magnetic field, or a larger amplitude turbulent velocities might be able to 
provide additional magnetic or turbulence support, encouraging high mass core
formation\footnote{However, the magnetized simulations of \citet{CO2014} show
no difference in median core masses from unmagnetized models}. 
Indeed, in
our high-amplitude perturbation tests, 3 among the 21 cores are very
massive with $ M_\mathrm{core}/M_0 \gtrsim 0.2$. These high-mass cores are 
all turbulent supported with $E_{k}/E_{th} \approx 2$ (most of the
smaller cores have $E_k/E_{th} < 1$).
This is suggestive that massive turbulent supported cores may be more prevalent
in systems with high velocity dispersion, but a more systematic study using
global rather than local models is required.

Alternatively, the final mass of a star may not be directly set by the mass
of the associated $t_\mathrm{coll}$-core. For example, 
more massive cores may be able to hold onto a larger fraction of 
their $t_\mathrm{coll}$-core mass, or capture more mass via competitive
accretion at late times from the mutual reservoir.
Those factors cannot be directly assessed in our simulations, since the sink
particles are artificially merged, and feedback processes are not
included. However, there is some evidence that the sink
particles in higher mass cores accrete faster, 
which is discussed in \S \ref{section:accretion}.

\subsection{Sink Particles}\label{section:sink}
\subsubsection{Accretion Rates of Sink Particles}\label{section:accretion}

\begin{figure*}[htbp]
\centering
\includegraphics[width=0.9\linewidth]{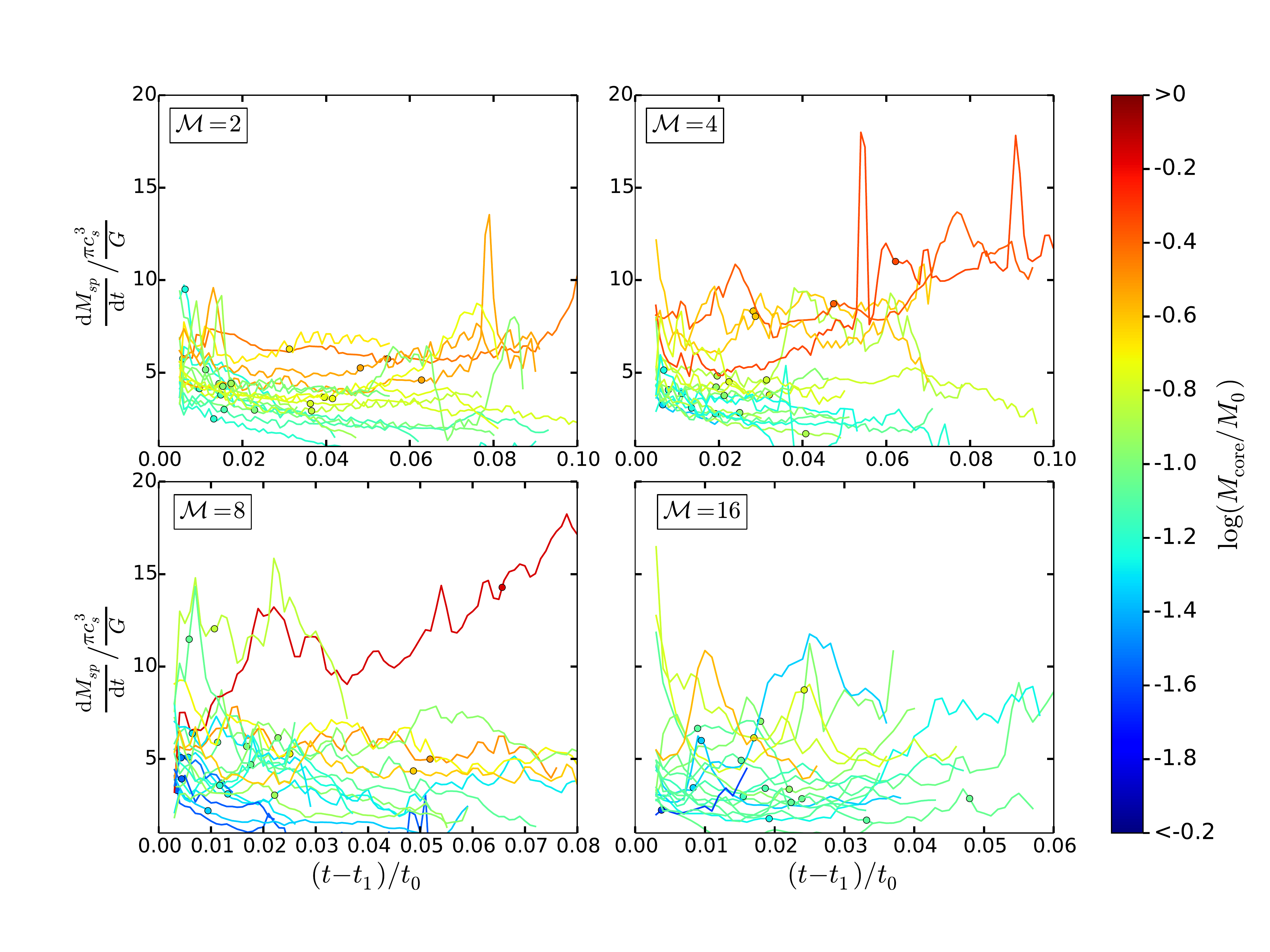}
\caption{Accretion rates of sink particles for different Mach numbers. We draw
a random subset of 40 sink particles in each Mach number case for demonstration.
Since mass in the control volume around a sink is added to the sink
instantaneously at the time of its creation, we plotted from time
$t_\mathrm{coll} + t_{ff}(\rho_{LP}(3\Delta x))$, approximately the time when
the mass within two times the radius of the control volume has fallen into the
sink, until one sink particle merges with another, or the end of the simulation. 
The accretion rates are calculated using the
slope of local cubic spline interpolation of the sink mass at different times,
to smooth out noise from discrete time intervals
(the time resolution of simulation output is $0.001t_0$).
The color of lines show the
corresponding $t_\mathrm{coll}$-core mass. 
The sphere on each line marks the time when the sink mass equals the
$t_\mathrm{coll}$-core mass. }
\label{fig:Mdot_t}
\end{figure*} 

When a sink particle is created, gravity dominates over gas
pressure in the envelope, and the gas accelerates into the sink, approaching free-fall.
A spherical LP density profile with gas free-falling to the 
center would have a constant mass accretion rate\footnote{The accretion rate
    unit, $M_0/t_0=\pi c_s^3/G$, is equal to $5.1\times
10^{-6}M_\sun\mathrm{yr}^{-1}$, for $c_s=0.19$km/s.}:
\begin{equation}
    \frac{\di(M_{sp}/M_0)}{\di (t/t_0)} = 7.6.
\end{equation}
In fact, pressure reduces the accretion rate somewhat (see below), but infall
still progresses from the inside to the outside of the envelope.
After this inside-out collapse reaches out to where the density profile
flattens out and there is more turbulent and thermal support, 
the accretion rate is expected to change.
Figure \ref{fig:Mdot_t} illustrates the accretion history of sink
particles. On average, the accretion rate remains roughly constant, with a
slight rising trend after the sink mass equals the $t_\mathrm{coll}$-core mass. 
However, we see no clear separation between the accretion of the initial
tidally-bound core and later stages; this is not surprising,
as the angle-averaged density profile is continuous
at the effective core radius. Noticeably, there are outbursts in accretion rates with
the peak several times higher than the average for a period of $\sim 0.01t_0$.
This is due to the clumpy density structures that develop and fall into the sink. 
In reality, the infall onto the protostar has to be processed through an
accretion disk which is not resolved in our simulations. 
The additional mass falling into the star during outbursts of this kind may
trigger gravitational or magnetic instabilities in the inner disk,
leading to more dramatic outbursts in YSOs, for instance, the episodic
accretion of FU Ori-type stars \citep{Audard2014, Ohtani2014}.

\begin{figure}[htbp]
\centering
\includegraphics[width=\linewidth]{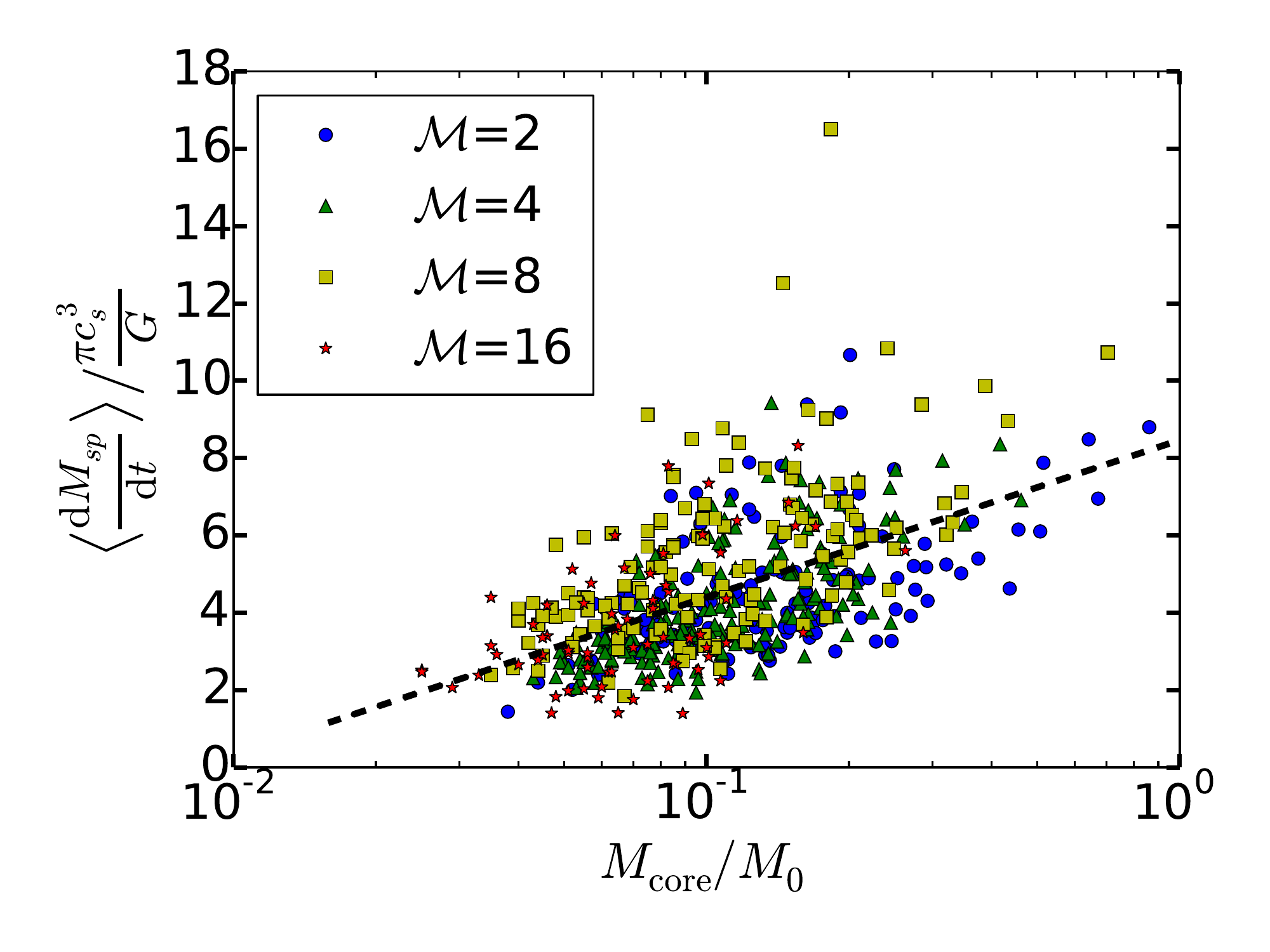}
\caption{Average accretion rates of sink particles from 
$t_\mathrm{coll} + t_{ff}(\rho_{LP}(3\Delta x))$ to the time when the sink mass
equals the corresponding $t_\mathrm{coll}$-core mass. The dotted
line shows a log-linear fit $ \di(M_{sp}/M_0)/\di (t/t_0) = 4.1\log
(M_\mathrm{core}/M_0) + 8.5$. With $c_s=0.19$km/s, The accretion rate unit is
$5.1\times 10^{-6} M_\sun\mathrm{yr}^{-1}$, such that most are in the range
$(1-5)\times 10^{-5} M_\sun\mathrm{yr}^{-1}$.}
\label{fig:Mdotavg}
\end{figure}

\begin{figure*}[htbp]
\centering
\includegraphics[width=0.8\linewidth]{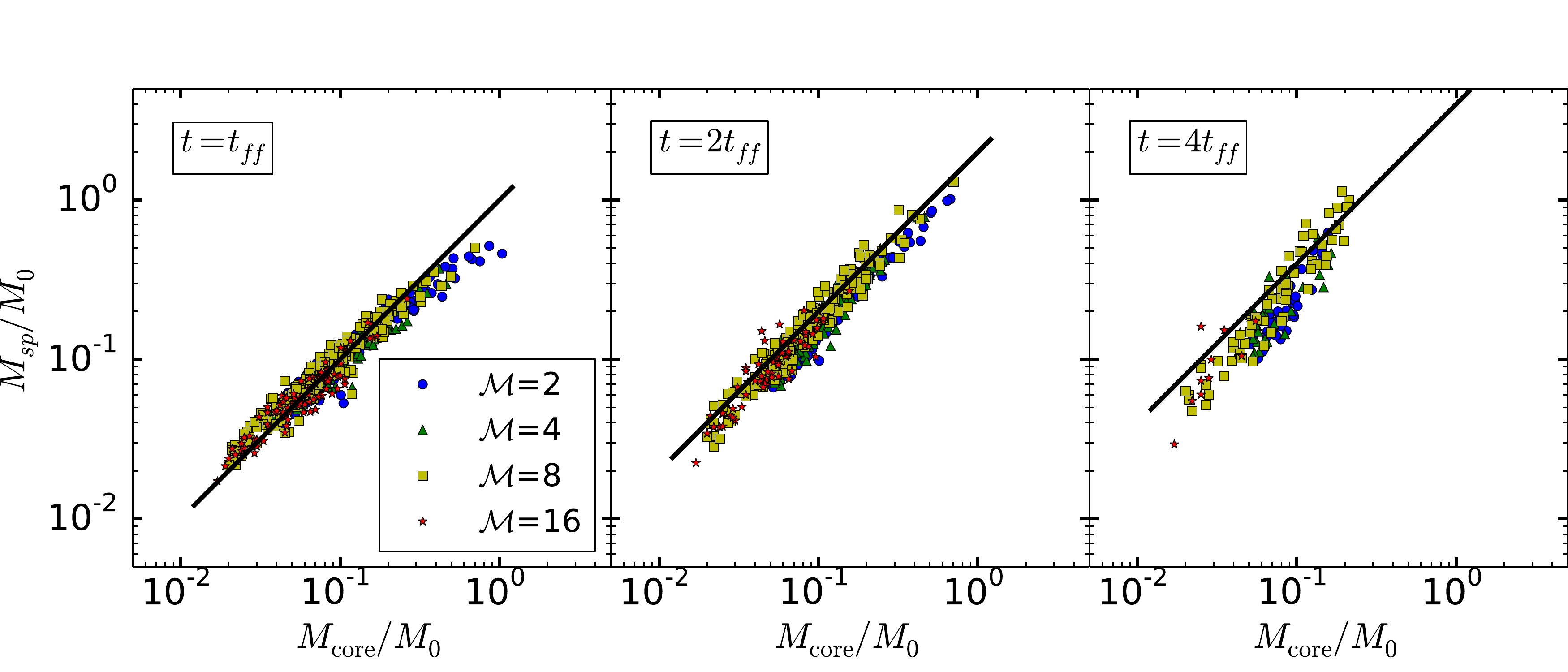}
\caption{The sink particle mass compared to the corresponding
    $t_\mathrm{coll}$-core mass
at 1, 2, and 4 times the free-fall time for each core. The
black lines plot $M_{sp}$ equals to 1, 2, and 4 times $M_\mathrm{core}$,
and are not fits to the data. Sinks accrete close to the $t_\mathrm{coll}$-core mass each
free-fall time, although high-mass cores begin to accrete faster than low-mass
cores over time (see right panel).}
\label{fig:Msp_Mcore_tff}
\end{figure*}

Figure \ref{fig:Mdot_t} also shows a trend that the average accretion rate
increases with the corresponding $t_\mathrm{coll}$-core mass,
which is quantified in Figure \ref{fig:Mdotavg}. 
The average $\dot{M}_{sp}$ depends linearly on $\log M_\mathrm{core}$, but 
there is also a big dispersion
with a factor of $\sim 2$ in accretion rates at a given core mass.
Another way to show the dependence of accretion rate on core mass is to look
at the mass of sink particle at the free-fall timescale of the core. Figure 
\ref{fig:Msp_Mcore_tff} shows sink particle masses at 1, 2, and 4 times the
initial free-fall time of the  $t_\mathrm{coll}$-cores. 
The initial mass in the core is accreted by the sink at $\sim t_{ff}$,
after which the accretion rate drops for low mass cores while it rises for high
mass ones. At $4t_{ff}$, the ratio of sink mass to $t_\mathrm{coll}$-core mass is a factor
of $\sim 2$ larger for the highest mass cores than the lowest mass ones. This
trend provides one possible explanation of the discrepancy between the CMF in our
simulations and the observed IMF, 
although in addition to this effect, the final stellar mass would depend on
feedback and mass loss at later stages of star formation, which is beyond the
scope of this paper.

The sink particle accretion properties of our high-amplitude velocity perturbation
simulations are similar to the low-amplitude cases discussed above.

\subsubsection{Star Formation Rate}\label{section:SFR}
\begin{figure*}[htbp]
\centering
\includegraphics[width=0.8\linewidth]{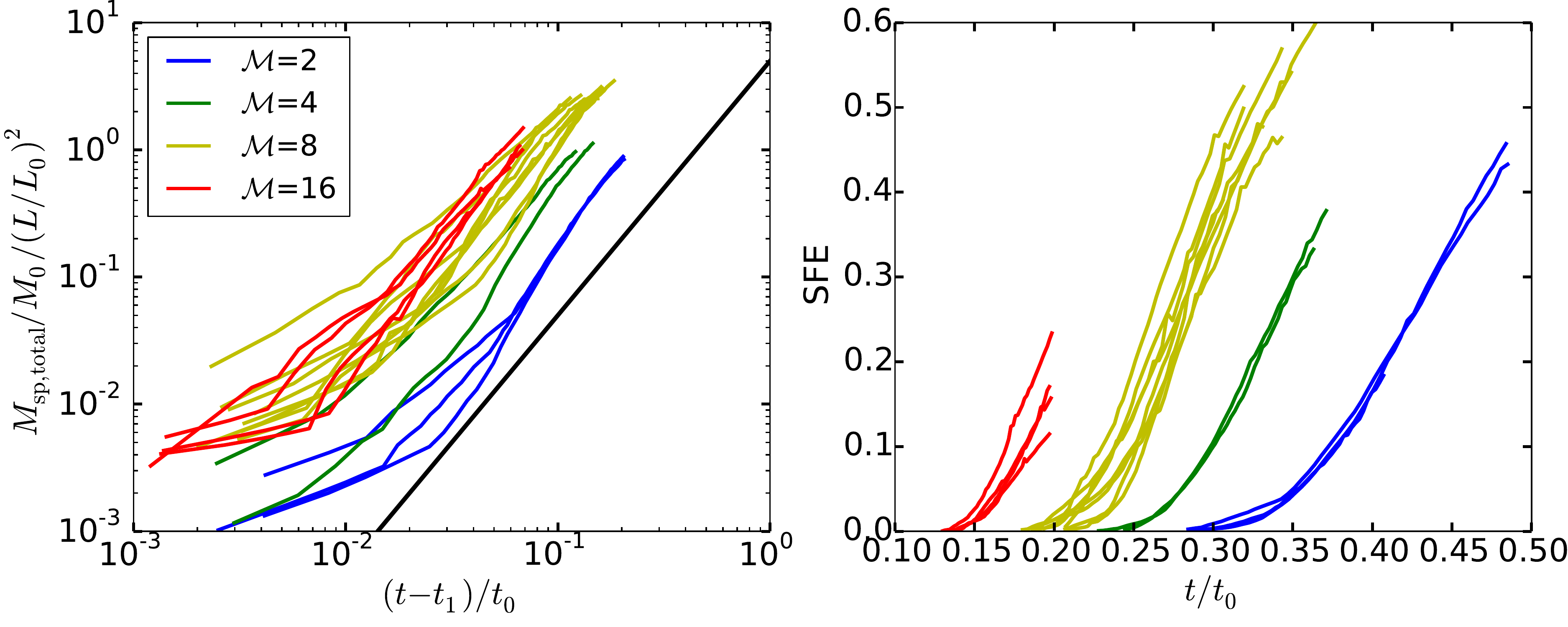}
\caption{Left: total mass in sink particles per unit area projected along the
$\hat{z}$ direction of the simulation box. Each line represents one simulation run. The
black solid line plots the slope of $M_\mathrm{sp,total}\propto t^2$. Right:
the star formation efficiency (SFE, see text for definition) of each simulation run.}
\label{fig:SFR}
\end{figure*} 

The total mass in sink particles, as shown in the left panel of 
Figure \ref{fig:SFR}, increases super-linearly with time, roughly as 
$M_\mathrm{sp,total}\propto t^2$. This is because both the 
mass of individual sinks and the total number of sink particles
increases linearly with time. \citet{Lee2014} also obtained a 
similar result that the accreted mass grows as
$t^2$ in turbulent MHD simulations.
For higher Mach numbers, we find that
there is more mass accreted at a given time,
both because sink particles form faster, and more sinks form per unit area. 
We note that this only represents the early stage of star formation when gravity
dominates, and the surrounding gas is abundant. The star formation rate would
have to drop at a later times when stellar feedback becomes important and
gas in the GMC is exhausted or removed.

The star formation efficiency (SFE) is shown in the
right panel of Figure \ref{fig:SFR}. The SFE is defined as the ratio of
total mass in all sinks to $2{\cal M}c_sL^2 t$, the total mass that has
entered the box (or the shocked gas). As the mass in sinks grows 
$\propto t^2$, the SFE
increases linearly with $t$. Similar to all simulations without
feedback (eg. \citet{Lee2014, Smith2009}), gas is accreted on the free-fall
timescale, resulting in a high SFE of $\sim 0.3-0.6$.

\section{Summary}\label{section:summary}
In this work, we have used three-dimensional numerical simulations to explore the
physics of pre-stellar core formation and the subsequent core collapse that leads to the
formation of protostars. We focus on local regions ($\sim 1$pc)
within GMCs where a supersonic turbulent flow converges, triggering star formation
in the post-shock layer. Our simulations adopt an idealized isothermal equation of
state, include gravity from both gas and sink particles, and explore a wide range
of Mach numbers (${\cal M}=2-16$) for the inflow gas.
We terminate our simulations at late
times, when most of the gas in the post-shock region has been accreted by the sink
particles.

For a fixed Mach number, we have carried out a series of
simulations with different resolutions and box sizes to study potential numerical
effects, especially the convergence of isothermal fragmentation. We have
obtained the properties of $\gtrsim 200$ cores in each model, for
both cores identified at the time of individual collapse
($t_\mathrm{coll}$-cores) and from a snapshot at the time of 
first singularity formation in each simulation ($t_1$-cores).
We study the statistical properties of cores,
and compare the CMFs in our simulations to the observed CMF and IMF. 
In addition, we trace the mass accretion rate of sink particles, 
and its dependence on the core mass.

Our main conclusions are as follows:\\
\begin{enumerate}
  \item Cores and self-gravitating filaments form and evolve at the same time.
      Isothermal filaments do not
      collapse to become singular spindles \citep[cf.][]{IM1992, IM1997}
      because both small-scale and large-scale over-densities are seeded by
      compression from turbulence, and filaments and cores grow simultaneously.

  \item At the time of singularity formation, all $t_\mathrm{coll}$-cores approach the
      Larson-Penston asymptotic solution, giving a mass-radius relation of $M
      \propto r$ (see Figure \ref{fig:mcore_rcore}),
      with the LP density profile extending beyond several times 
      the angle-averaged tidal radius of the cores. 
      Most $t_\mathrm{coll}$-cores are prolate, consistent with
      filament fragmentation. The mass-radius relation of $t_1$ cores implies a
      flatter density profile with $M \propto r^{1.2-2.5}$, similar to
      observations \citep[e.g.,][]{CR2010, Kirk2013}. Many $t_1$-cores are
      only weakly bound.

  \item At the stage of core collapse, there is a well-defined characteristic
      mass scale for fragmentation in our turbulent,
      self-gravitating simulations.
      The characteristic $t_\mathrm{coll}$-core mass in our simulations converges as
      we increase the resolution and density threshold for sink particle
      creation. This mass is very close to 
      the mass of a critical Bonner-Ebert sphere (see Equation
      (\ref{eq:M_BE_crit})), taking the ``edge" as equal to the mean pressure
      in the post-shock layer formed by the converging flow 
      (See Figure \ref{fig:mBE_Mach_mcore}).

  \item The characteristic core mass for the present simulations is insensitive
      to the Mach number of the large-scale converging flow.
      As the inflow Mach number increases, the post-shock density and pressure do not
      vary much, due to instabilities that disperse the gas in the post-shock
      layer, leading to similar core properties across different Mach numbers.
      For all models, the characteristic core mass is $\sim 1M_\sun$.
      We note, however, that inclusion of magnetic fields reduces these
      instabilities, such that the (magnetically-dominated)
      post-shock pressure is higher, and
      characteristic core mass is lower in models with higher Mach numbers
      (Chen \& Ostriker 2015, in preparation)\footnote{However, in the
      magnetized case, the characteristic core mass is still comparable to the
      critical Bonner-Ebert mass at the mean post-shock pressure}.
      
  \item The CMFs of $t_\mathrm{coll}$-cores in our simulations show a
      log-normal distribution with standard deviation $\sim 0.3-0.4$ (Figure
      \ref{fig:CMF_tcoll}). This is similar to the bound core distribution
      width for Aquila \citep{Konyves2010}. 
      There is a significant deficit of high mass cores ($\gtrsim 7M_\sun$) compared to
      the shape of the observed IMF. However, the CMFs of $t_1$-cores (Figure
      \ref{fig:CMF_t1}) show a wider
      distribution with more low mass cores, similar to the shape of
      the full observed CMF \citep[e.g.,][]{Alves2007, NW2007, Andre2010}.
      This suggests that observed CMFs, similar to our
      $t_1$-cores, may include cores at an early stage of evolution that will
      gain a significant amount of mass before collapsing, or transient structures
      from turbulence perturbations that may never collapse. 
      Therefore, the observed CMF from a cloud
      ``snapshot"
      may not represent the distribution of mass reservoirs for
      protostar collapse. We find, however, that the $t_1$-cores that exceed
      the critical BE mass have a distribution that is closer to that of the
      $t_\mathrm{coll}$-cores; a similar selection could be made for observed
      cores.

  \item The mass accretion rate of sink particles increases weakly with the
      corresponding $t_\mathrm{coll}$-core mass (Figure \ref{fig:Msp_Mcore_tff}), 
      suggesting that competitive
      accretion may play a role in filling out the high-mass portion of the
      IMF. 
      Individual sink particles accrete at roughly constant $\dot{M}_{sp}$, 
      and continue to gain more mass even after they have exceeded the mass of
      the core when it first collapsed.
      There are outbursts in  $\dot{M}_{sp}\approx (1-5)\times 10^{-5}
      M_\sun\mathrm{yr}^{-1}$ (Figures \ref{fig:Mdot_t}, \ref{fig:Mdotavg}) 
      due to clumpy density structures forming and falling into the sink.

  \item The star formation rate (measured as the total mass in sink particles)
      in each converging-flow simulation 
      increases with time as $\mathrm{SFR} \propto t$, as both the formation rate of
      sink particles (i.e., rate of cores surpassing the collapse threshold),
      and the accretion rate of individual sinks are roughly constant.
      This result is similar to the ``accelerating star formation" found in some
      other recent simulations \citep{Lee2014}. For a realistic cloud, however,
      the duration of any given large-scale converging flow would be less than
      the crossing time of the cloud, and feedback would limit the accretion
      within any individual core.
\end{enumerate}

The present suite of models extend previous simulation studies of
gravoturbulent fragmentation in several important ways. Most notably, by
systematically varying the Mach number, we are able to test the influence of a
key environmental parameter, and by systematically varying the resolution, we
are able to demonstrate that evolution robustly includes a core-collapse
stage with well-defined mass. The physical ingredients of our simulations are
limited, however, in particular lacking magnetic fields, feedback from forming
stars, and effects of environmental evolution. Furthermore, global cloud
simulations are required in order to incorporate the full spectrum of
turbulence at realistic amplitudes.
Other work currently underway
will address these limitations while maintaining a focus on systematic surveys of
the environmental parameter space.

\acknowledgments
Acknowledgements: This work was supported in part by grant NNX10AF60G from the NASA Astrophysics
Theory Program. We are grateful to the referee for a helpful report.

\bibliographystyle{apj}
\bibliography{apj-jour,coreform}
\end{document}